\documentclass[aps,prd]{revtex4}
\usepackage{epsfig,epsf}
\usepackage{amsmath}
\usepackage{amsthm}
\usepackage{amsfonts}
\usepackage{amssymb}
\usepackage{dsfont}
\usepackage{multirow}
\usepackage{appendix}
\usepackage{slashed}
\usepackage[active]{srcltx}
\usepackage{psfrag}
\usepackage{subfigure}
\usepackage[colorlinks,citecolor=blue,urlcolor=red,linkcolor=red]{hyperref}

\begin{document}
\title{{\Large{\bf Semileptonic $B_{(s)}\to a_1(K_1)\ell^+ \ell^-$ decays via the light-cone sum rules with $B$-meson distribution amplitudes}}}

\author{S. Momeni}%
\email[]{e-mail: samira.momeni@phy.iut.ac.ir}
\author{R. Khosravi}%
\email[]{e-mail: rezakhosravi @ cc.iut.ac.ir}

\affiliation{Department of Physics, Isfahan University of
Technology, Isfahan 84156-83111, Iran}

\begin{abstract}
The form factors of  semileptonic $B_{(s)}\to a_{1}(K_1)\, \ell^{+}
\ell^{-}$, $ \ell=\tau,\mu,e$ transitions are investigated  in the
framework of the light-cone sum rules with $B$-meson distribution
amplitudes, which play an important role in  exclusive $B$ decays.
The $B$-meson distribution amplitudes, $\varphi_{\pm}(\omega)$ are a
model-dependent form, so we consider four different
parameterizations which can provide a reasonable description of
$\varphi_{\pm}(\omega)$ from QCD corrections. The branching
fractions of these transitions  are calculated. For a better
analysis, a comparison of our results with the prediction of other
models is provided.
\end{abstract}

\maketitle

\section{Introduction}
Inclusive and exclusive decays of $B$-meson play a perfect role in
determination of fundamental parameters used in the standard model
(SM) and improve our studies in understanding the dynamics of
quantum chromo dynamics (QCD). Among of all $B$ decays, the
semileptonic decays occupy a special place  since their theoretical
description is relatively simple. In this field, reliable
calculations of heavy-to-light transition form factors of
semileptonic $B$ decays are very important in particle physics.
These form factors are also used to determine the amplitude of
non-leptonic $B$ decays  applied to evaluate the CKM parameters as
well as to test  various properties of the SM.

In the region of large momentum transfer squared, ($q^2$)
heavy-to-light form factors are successfully investigated via the
Lattice QCD. But in small $q^2$, other approaches are used such as
the light-cone sum rules (LCSR)
\cite{Kolesnichenko,Halperin,Zhitnitsky}. In the usual LCSR method,
the correlation function is inserted between the vacuum and light
meson. As a result of this calculation, the long distance dynamics
is described by light-cone distribution amplitudes (LCDA's) of light
meson
\cite{Ruckl,Simma,Belyaev,Weinzierl,Bagan,Ball,Zwicky,BZwicky,BBraun}.
Still, there is  very limited knowledge  of the nonperturbative
parameters determining these LCDA's. Therefore, the main uncertainty
in estimating  the form factors comes from the limited accuracy of
the  LCDA parameters.

As the direct analogue of the LCDA's of light mesons, the $B$-meson
distribution amplitudes (DA's) were introduced to describe generic
exclusive $B$ decays with the contribution of the hard gluon
exchange \cite{Brodsky}. Based on the local OPE and condensate
expansion, the classical two-point sum rules was used for the
$B$-meson DA's already in the original study \cite{Grozin}. The
$B$-meson DA's emerge as universal nonperturbative objects in many
studies of exclusive $B$-meson decays (for instance see
\cite{Beneke}). An estimate of the inverse moment of two-particle
DA, $\varphi_{+}$ was also obtained by matching the factorization
formula to the LCSR for $B\to \gamma \ell \nu$ \cite{Kou}. The
shapes of the $B$-meson DA, $\varphi_{+}$ depends on the model and
our knowledge of the behaviors of $\varphi_{+}$ is still rather
limited due to the poor understanding of nonperturbative QCD
dynamics.

Using the LCSR technique and relating the $B$-meson DA's to the
$B\to \pi $  form factor, a new approach was suggested in
\cite{Offen}. In this new LCSR, correlation function was taken
between the vacuum and $B$-meson and it was expanded  in terms of
$B$-meson DA's near the light-cone region. Therefore, the link was
established between the $B$-meson DA's and transition form factors,
which  provide an independent dynamical information on the $B$-meson
DA's. The new LCSR has been derived for $B\to \pi,K$ and $B \to
\rho,K^*$ form factors in the leading order including the
contributions of two- and three-particle DA's in \cite{Mannel}.
Moreover, in this reference the $B$-meson three-particle DA's have
been investigated and their form  have been established at small
momenta of light-quark and gluon.

In this paper the  heavy-to-light decays, $B_{(s)}\to a_{1}(K_1)\,
\ell^{+} \ell^{-}, \ell = e, \mu, \tau$  are described by the flavor
changing neutral current (FCNC) processes via $b\to d\,\ell^{+}
\ell^{-}$ transition at quark level which proceed through the
electroweak penguin and box diagrams. The exclusive FCNC $B$ decays
are important for development of new physics and flavor physics
beyond the SM. The main purpose of this paper is to consider the
form factors of the FCNC $B_{(s)} \to a_1(K_1)$ transition with LCSR
approach, using the $B$-meson DA's, and comparing these  form
factors with those of other approaches, especially the usual LCSR.
Comparing form factor results between two independent methods
establish input parameters and assumptions as well as  predictions
of the conventional LCSR.

It should be noted that the physical state of $K_1(1270)$ meson is
consider as a mixture of two $|^3P_1\rangle$ and $|^1P_1\rangle$
states and can be parameterized in terms of a mixing angle
$\theta_{K}$, as follows \cite{Hatanaka}:
\begin{eqnarray}\label{eq:mixing}
|K_{1}(1270)\rangle &=& \sin\theta_{K} |^3P_{1}\rangle ~+
\cos\theta_{K} |^1P_{1}\rangle,
\end{eqnarray}
where $|^3P_1\rangle\equiv |K_{1A}\rangle$ and $|^1P_1\rangle\equiv
|K_{1B}\rangle$ with different masses and decay constants. Also
$\theta_K$ is the mixing angle and can be determined by the
experimental data. There are various approaches to estimate the
mixing angle. In \cite{Burakovsky} the result $35^\circ< |\theta_K|
< 55^\circ$ was found  while in   ~\cite{Suzuki}, two possible
solutions with   $|\theta_K|\approx 33^\circ$ and $57^\circ$ were
obtained.

The contents of this paper are as follows: In section II, the
effective weak Hamiltonian of the $b \to d~ \ell^+ \ell^-$
transition are presented. In section III, we derive the $B_{(s)}\to
a_{1}(K_1)\, \ell^{+} \ell^{-}$ form factors with the LCSR method
using the $B$-meson DA's. To achieve a better analysis, we consider
four different parameterizations for the shapes of the $B$-meson
DA's, $\varphi_{\pm}$. The form factors of the $B_{(s)}\to
a_{1}(K_1)\, \ell^{+} \ell^{-}$ decays are basic parameters in
studying the exclusive non-leptonic two-body decays and semileptonic
decays. Our numerical analysis of the form factors as well as
branching ratio values and their comparison with the prediction of
other approaches is provided in section IV.

\section{The effective weak Hamiltonian of the $b \to
d~ \ell^+ \ell^-$ transition}
In the SM, the  $B_{(s)}\to
a_{1}(K_1)\, \ell^+ \ell^-$ decay amplitude is reduced to the matrix
element defined as $\langle a_1(K_1)\, \ell^+ \ell^- |\mathcal
{H}^{b\to d}_{\rm eff}|B_{(s)}\rangle. $ The effective weak
Hamiltonian of the $b \to d~ \ell^+ \ell^-$ transition  has the
following form in the SM:
\begin{eqnarray}\label{eq21}
H^{b\to d}_{\rm eff} = - \frac{ G_F}{\sqrt{2}}\left(
V_{ub}V_{ud}^{*}\sum_{i=1}^{2} C_i(\mu)
O^{u}_i(\mu)+V_{cb}V_{cd}^{*}\sum_{i=1}^{2} C_i(\mu)
O^{c}_i(\mu)-V_{tb}V_{td}^{*}\sum_{i=3}^{10} C_i(\mu)
O_i(\mu)\right),
\end{eqnarray}
where $V_{jk}$ and $C_i(\mu)$ are the  CKM matrix elements and
Wilson coefficients, respectively. The local operators are
current-current operators $O^{u,c}_{1,2}$, QCD penguin operators
$O_{3-6}$, magnetic penguin operators $O_{7,8}$, and semileptonic
electroweak penguin operators $O_{9,10}$. The explicit expressions
of these operators  for $b \to d \ell^+ \ell^-$ transition are
written as \cite{Buras0}
\begin{eqnarray}
\begin{array}{ll}
O_1     =  (\bar{d}_{i}  c_{j})_{V-A} ,
           (\bar{c}_{j}  b_{i})_{V-A}  ,               &
O_2     =  (\bar{d} c)_{V-A}  (\bar{c} b)_{V-A}   ,              \\
O_3     =  (\bar{d} b)_{V-A}\sum_q(\bar{q}q)_{V-A}  ,            &
O_4     =  (\bar{d}_{i}  b_{j})_{V-A} \sum_q (\bar{q}_{j}
          q_{i})_{V-A} ,                                    \\
O_5     =  (\bar{d} b)_{V-A}\sum_q(\bar{q}q)_{V+A} ,            &
O_6     =  (\bar{d}_{i}  b_{j })_{V-A}
   \sum_q  (\bar{q}_{j}  q_{i})_{V+A}  ,               \\
O_7     =  \frac{e}{8\pi^2} m_b (\bar{d} \sigma^{\mu\nu}
          (1+\gamma_5) b) F_{\mu\nu}  ,                     &
O_8    =  \frac{g}{8\pi^2} m_b (\bar{d}_i \sigma^{\mu\nu}
   (1+\gamma_5) { T}_{ij} b_j) { G}_{\mu\nu}  ,          \\
O_9     = \frac{e}{8\pi^2} (\bar{d} b)_{V-A}  (\bar{l}l)_V ,      &
O_{10}  = \frac{e}{8\pi^2} (\bar{d} b)_{V-A}  (\bar{l}l)_A ,
\end{array}
\end{eqnarray}
where ${ G}_{\mu\nu}$ and $F_{\mu\nu}$ are the gluon and photon
field strengths, respectively; ${T}_{ij}$ are the generators of
the $SU(3)$ color group; $i$ and $j$ denote color indices.
Labels $(V\pm A)$ stand for
$\gamma^\mu(1\pm\gamma^5)$.
The magnetic and
electroweak penguin operators $O_{7}$, and $O_{9,10}$ are
responsible for the short distance (SD) effects in the FCNC $b \to
d$ transition, but  the operators $O_{1-6}$  involve both SD and
long distance (LD) contributions in this transition.
In the naive
factorization approximation, contributions of the $O_{1-6}$
operators have the same form factor dependence as $C_9$ which can be
absorbed into an effective Wilson coefficient $C^{\rm eff}_9$.
Therefore, the matrix element for the $b \rightarrow d \ell^+
\ell^-$ transition can be written as:
\begin{eqnarray}\label{eq27}
\cal{M}&=& \frac{G_{F}\alpha}{2\sqrt{2}\pi} V_{tb}V_{td}^{*}\Bigg[
C_9^{\rm eff}  \overline {d} \gamma_\mu (1-\gamma_5) b~  \overline
{l }\gamma_\mu l + C_{10} \overline {d} \gamma_\mu (1-\gamma_5) b~
\overline {l } \gamma_\mu \gamma_5 \l \nonumber\\&-& 2 C_7^{\rm
eff}\frac{m_b}{q^2} \overline {d} ~i\sigma_{\mu\nu} q^\nu
(1+\gamma_5) b~  \overline {l}  \gamma_\mu l \Bigg],
\end{eqnarray}
where $\bar d \gamma_\mu (1-\gamma_5) b$ and $\bar d\, i
\sigma_{\mu\nu}q^\nu (1+\gamma_5) b$ are the transition currents
denoted with $J^{V-A}_{\mu}$ and $J^{T}_{\mu}$ respectively, in this
work. Eq. (\ref{eq27}) also contains two effective Wilson
coefficients $C_7^{\rm eff}$ and $C_9^{\rm eff}$, where $C_7^{\rm
eff}= C_7-C_5/3-C_6$. The effective Wilson coefficient
$C_{9}^{\rm{eff}}$ includes both the SD and LD effects as
\begin{eqnarray}\label{eq22}
C^{\rm eff}_9 = C_9 + Y_{SD}(q^2)+Y_{LD}(q^2),
\end{eqnarray}
where $Y_{SD}(q^2)$ describes the SD contributions from four-quark
operators far away from the resonance regions, which can be
calculated reliably in perturbative theory as \cite{Buras0,Aliev0}:
\begin{eqnarray}\label{eq23}
Y_{SD}(q^2)&=&0.138~ \omega(s)+h(\hat{m_c},s)C_0
-\frac{1}{2}h(1,s)(4C_3+4C_4+3C_5+C_6)\nonumber\\
&-&\frac{1}{2}h(0,s)(2\lambda_u [3C_1+C_2]+C_3+4C_4)
+\frac{2}{9}(3C_3+C_4+3C_5+C_6),
\end{eqnarray}
where $s=q^2/m_b^2$, $\hat{m_c}=m_c/m_b$,
$C_0=-\lambda_c(3C_1+C_2)+3C_3+C_4+3C_5+C_6$, $\lambda_c
=\frac{V_{cb}V^*_{cd}}{V_{tb}V^*_{td}}$, $\lambda_u
=\frac{V_{ub}V^*_{ud}}{V_{tb}V^*_{td}}$,    and
\begin{eqnarray}\label{eq024}
\omega(s)&=& -\frac{2}{9} \pi^2 -\frac{4}{3} {\rm Li}_2(s)-
\frac{2}{3} \ln (s) \ln(1-s) - \frac{5+4s}{3(1+2s)} \ln(1-s) -
\frac{2 s(1+s)(1-2s)}{3(1-s)^2(1+2s)} \ln (s)\nonumber \\ &+&
\frac{5+9 s-6 s^2}{3(1-s)(1+2s)},
\end{eqnarray}
represents the ${\cal O}(\alpha_s)$ correction coming from one gluon
exchange in the matrix element of the operator $O_9$ \cite{Jezabek},
while  $h(\hat m_c,  s)$ and $h(0,  s)$ represent one-loop
corrections to the four-quark operators $O_{1-6}$ \cite{Misiak}. The
functional form of the $h(\hat{m_c}, s)$ and $h(0, s)$ are as:
\begin{eqnarray}\label{eq025}
h(\hat m_c,  s) & = & - \frac{8}{9}\ln\frac{m_b}{\mu} -
\frac{8}{9}\ln\hat m_c +
\frac{8}{27} + \frac{4}{9} x \nonumber\\
& - & \frac{2}{9} (2+x) |1-x|^{1/2} \left\{
\begin{array}{ll}
\left( \ln\left| \frac{\sqrt{1-x} + 1}{\sqrt{1-x} - 1}\right| - i\pi
\right), &
\mbox{for } x \equiv \frac{4 \hat m_c^2}{ s} < 1 \nonumber \\
 & \\
2 \arctan \frac{1}{\sqrt{x-1}}, & \mbox{for } x \equiv \frac {4 \hat
m_c^2}{ s} > 1,
\end{array}
\right. \nonumber\\
h(0,  s) & = & \frac{8}{27} -\frac{8}{9} \ln\frac{m_b}{\mu} -
\frac{4}{9} \ln\ s + \frac{4}{9} i\pi.
\end{eqnarray}
The LD contributions, $Y_{LD}(q^2)$ from four-quark operators near
the $u\bar{u}$, $d\bar{d}$ and $c\bar{c}$ resonances can not be
calculated from the first principles of QCD and are usually
parameterized in the form of a phenomenological Breit-Wigner formula
as \cite{Buras0,Aliev0}:
\begin{eqnarray}\label{eq026}
Y_{LD}(q^2)&=&\frac{3\pi}{\alpha^2}
\left\{\sum_{V_i=\psi(1s),\psi(2s)}\frac{\Gamma(V_i\to l^+
l^-)m_{V_i}}{m_{V_i}^2-q^2-i m_{V_i} \Gamma_{V_i}} - \lambda_u
h(0,s)(3C_1+C_2)\sum_{V_i=\rho,\omega}\frac{\Gamma(V_i\to l^+
l^-)m_{V_i}}{m_{V_i}^2-q^2-i m_{V_i} \Gamma_{V_i}}\right\}.\nonumber\\
\end{eqnarray}

\section{$B_{(s)}\to a_{1}(K_1)\, \ell^{+} \ell^{-}$  form factors with the LCSR}
\begin{figure}[th]
\includegraphics[width=4cm,height=3cm]{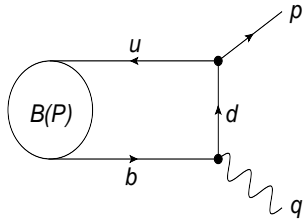}
\caption{leading-order diagram for $B \rightarrow
a_1 \ell^+ \ell^-$  decays.} \label{F22}
\end{figure}

First, we start with the two-point correlation function to compute
the form factors of the $B\to a_{1} \ell^{+} \ell^{-}$ via the LCSR
and then explain how to extract the $B\to K_{1} $ transition form
factors. The correlation function is constructed from the transition
currents $J^{V-A}_{\nu}$ and $J^{T}_{\nu}$ as follows:
\begin{eqnarray} \label{eq28}
\Pi_{\mu\nu}^{V-A~(T)}(p,q)=i\int {d^{4}x}~ e^{ip.x}
\langle0|\mathcal T\left\{
J_{\mu}^{a_1}(x)J_{\nu}^{V-A~(T)}(0)\right\}|B(P)\rangle.
\end{eqnarray}
In this definition for  the correlation function, $\mathcal T$ is
the time ordering operator, $|0\rangle$  is an appropriate ground
state (usually vacuum), $J_{\mu}^{a_1}=\bar{u}\gamma_{\mu}\gamma_{5}
d $ is the interpolating current of the axial-vector meson $ a_{1}
$. The external momenta of the interpolating and transition
currents, $J_{\mu}^{a_1}$ and $J_{\nu}^{V-A~(T)}$, are $p$ and $q$,
respectively, and $P^2= (p + q)^2 = m^2_B$. The leading-order
diagram for $B \rightarrow a_1 \ell^+ \ell^-$  decays is depicted in
Fig. \ref{F22}.

According to the general philosophy of the QCD sum rules and its
extension, (light-cone sum rules), the above correlation function
should be calculated in two different ways. In phenomenological or
physical representation, it is calculated in terms of hadronic
parameters. In QCD side, it is obtained in terms of DA's and QCD
degrees of freedom. The LCSR for the physical quantities like form
factors are acquired equating coefficient of the sufficient
structures from both representations of the same correlation
function through the dispersion relation and applying Borel
transformation and continuum subtraction to suppress the
contributions of the higher states and continuum.

To obtain the phenomenological representation of the correlation
function, a complete set of intermediate states with the same
quantum number as the current $J^{a_1}_{\mu}$ is inserted in Eq.
(\ref{eq28}). Isolating the pole term of the lowest axial vector
$a_1$ meson and applying Fourier transformation, we get
\begin{eqnarray}\label{eq29}
\Pi^{V-A~(T)}_{\mu\nu} (p,q) &=& \frac{1}{p^{2}-m_{a_1}^2}
\langle 0 | J^{a_1}_{\mu}(p) | a_1(p) \rangle \langle a_1(p) | J^{V-A~(T)}_{\nu}| B(P) \rangle
+ \mbox{higher states\,.}
\end{eqnarray}
The matrix element, $\langle 0 | J^{a_1}_{\mu}(p) | a_1(p) \rangle$
is defined as
\begin{eqnarray} \label{eq210}
\langle 0|J_{\mu}^{a_1}(0)|a_{1}(p)\rangle =f_{a_1}m_{a_1}\varepsilon_{\mu},
\end{eqnarray}
where $f_{a_1}$ and  $ \varepsilon _{\mu}$  are  the leptonic decay
constant and  polarization vector of the axial vector meson $a_1$,
respectively. The transition matrix element, $\langle a_1(p) |
J^{V-A~(T)}_{\nu}| B(P) \rangle$, can be parameterized via Lorentz
invariance and parity considerations as \cite{Colangelo}:
\begin{eqnarray}\label{eq211}
\langle a_{1}(p)|J_{\mu}^{V-A} | B(P)\rangle&=& \epsilon_{\mu\nu\alpha\beta} \varepsilon^{* \nu} P^{\alpha} p^{\beta} \;  \frac{2A(q^{2})}{m_{B}-m_{a_1}}
-~i\varepsilon^{\ast}_{\mu}(m_{B}-m_{a_1})V_{1}(q^{2}) \nonumber\\
&+&i \frac{\varepsilon^{\ast}\cdot P}{m_{B}-m_{a_1}}(P+p)_{\mu} V_{2}(q^{2}) +~ 2 i m_{a_1}\frac{\varepsilon^{\ast}\cdot P} {q^{2}} q_{\mu}[V_{3}(q^{2})-V_{0}(q^{2})], \nonumber \\
\langle a_{1}(p)|J_{\mu}^{T}  |B(P) \rangle&=& 2 i \epsilon_{\mu \nu \alpha \beta} \varepsilon^{* \nu} P^{\alpha}p^{\beta}
\; T_1(q^2)+[\varepsilon^*_\mu (m_B^2 - m^2_{a_1})-~(\varepsilon^* \cdot P) (P+p)_\mu ] \; T_2(q^2) \nonumber \\
&+& (\varepsilon^* \cdot P) [ q_\mu - {q^2 \over m_B^2 - m^2_{a_1}} (P+p)_\mu]\;  T_3(q^2),
\end{eqnarray}
$ q^{2}$ is the momentum transfer squared of the $Z$  boson
(photon). It should be noted that $V_{0}(0)=V_{3}(0) $ and the
identity
$\sigma_{\mu\nu}\gamma_{5}=\frac{-i}{2}\epsilon_{\mu\nu\alpha\beta}\sigma^{\alpha\beta}
(\epsilon_{0123}=1)  $    implies that $T_{1}(0)=T_{2}(0)
$\cite{Colangelo}. Moreover, $V_{3} $  can be written as a linear
combination of  $V_{1}  $ and $V_{2} $:
\begin{eqnarray}\label{eq212}
V_{3}(q^{2})=\frac{m_{B}-m_{a_1}}{2m_{a_1}} V_{1}(q^{2})-\frac{m_{B}+m_{a_1}}{2m_{a_1}} V_{2}(q^{2}).
\end{eqnarray}

Using Eqs. (\ref{eq210}) and (\ref{eq211})  in Eq. (\ref{eq29}), and
performing summation over the polarization of $a_1$ meson, we obtain
\begin{eqnarray}\label{eq213}
\Pi_{\mu\nu}^{V-A}&=&\frac{f_{a_1}m_{a_1}}
{p^2-m_{a_1}^2} \times \left[ \frac{2 {A}}{m_{B}-m_{a_1}}(q^2)\,\epsilon_{\mu\nu\alpha\beta}P^{\alpha}p^{\beta}
-i~{V}_{1}(q^2)\,(m_{B}-m_{a_1})\,g_{\mu\nu}\right.\nonumber\\&+i&\left.\frac{{V}_{2}(q^2)}{m_{B}-m_{a_1}}\, (P+p)_{\mu}P_{\nu} -i~\frac{2m_{a_1}{V}_{0}(q^2)}{q^2}\,q_{\mu}P_{\nu}\right]
+ \mbox{higher states}\,,\nonumber\\
\Pi_{\mu\nu}^{T}&=&\frac{f_{a_1}m_{a_1}}
{p^2-m_{a_1}^2} \times \left[2\,{T
}_{1}(q^2)\,\epsilon_{\mu\nu\alpha\beta}P^{\alpha}p^{\beta}
-i~ {T}_{2}(q^2)\,(m_B^2-m_{a_1}^{2})\,g_{\mu\nu}\right.\nonumber\\&-i&\left. {T}_{3}(q^2)\,q_{\mu}P_{\nu}
\right]+ \mbox{higher states}\,.
\end{eqnarray}

To calculate the form factors $A$, $V_i (i=0, 1, 2)$ and ${T}_{j}
(j= 1, 2, 3)$, we will choose the structures
$\epsilon_{\mu\nu\alpha\beta} P^{\alpha} p^{\beta}$, $g_{\mu\nu}$,
$(P+p)_{\mu}P_{\nu}$, $q_{\mu}P_{\nu}$, from $\Pi_{\mu\nu}^{V-A}$
and $\epsilon_{\mu\nu\alpha\beta} P^{\alpha}p^{\beta}$,
$g_{\mu\nu}$, and $q_{\mu}P_{\nu}$ from $\Pi_{\mu\nu}^{T}$,
respectively. For simplicity, the correlations are written as
\begin{eqnarray}\label{eq214}
\Pi_{\mu\nu}^{V-A}(p,q)&=&\Pi_{1}~g_{\mu\nu}+\Pi_{2}~\epsilon_{\mu\nu\alpha\beta} P^{\alpha} p^{\beta}+\Pi_{3}~ (P+p)_{\mu}P_{\nu}+\Pi_{4}~q_{\mu}P_{\nu}+.....,\nonumber\\
\Pi_{\mu\nu}^{T}(p,q)&=&\Pi_{1}^{'}~g_{\mu\nu}+\Pi_{2}^{'}~\epsilon_{\mu\nu\alpha\beta} P^{\alpha}p^{\beta}+ \Pi_{3}^{'}~q_{\mu}P_{\nu}+...~.
\end{eqnarray}

Now, we consider the QCD part of the correlation functions in Eq.
(\ref{eq28}) based on light-cone OPE in the heavy quark effective
theory (HQET). After the transition to HQET, the correlation
functions are written as \cite{Mannel}:
\begin{eqnarray}\label{eq215}
\Pi_{\mu\nu}^{V-A (T)}(p,q)&=&\widetilde{\Pi}_{\mu\nu}^{V-A
(T)}(p,\widetilde{q})+\mathcal{O}(1/m_b),
\end{eqnarray}
where $\widetilde{q}=q-m_b v$, and $v$ is the four-velocity of
$B$-meson. Also up to $1/m_b$ corrections in HQET,  the state of
$B$-meson $|B(P)\rangle$,  and the $b$-quark field $b(x)$ are
substituted by the state $|B(v)\rangle$ and the effective field
$e^{-im_bvx} h_v(x)$, respectively. Therefore, the correlation
functions in the heavy quark limit, ($m_b \to \infty$), become:
\begin{eqnarray}\label{eq216}
\widetilde{\Pi}_{\mu\nu}^{V-A }(p,\widetilde{q})&=& i\int d^4x~
e^{ip.x}\langle 0|\mathcal T \{\bar{u}(x)\gamma_{\mu}\gamma_{5}~
iS_d(x)~ \gamma_{\nu} (1- \gamma_{5}) h_v(0)\}
|B(v)\rangle,\nonumber\\
\widetilde{\Pi}_{\mu\nu}^{T}(p,\widetilde{q})&=& i\int d^4x~
e^{ip.x}\langle 0|\mathcal T \{\bar{u}(x)\gamma_{\mu}\gamma_{5}~
iS_d(x)~   \sigma_{\nu\eta}q^\eta (1+\gamma_5)  h_v(0)\}
|B(v)\rangle.
\end{eqnarray}
From Eq. (\ref{eq216}) a convolution of a short-distance part with
the matrix element of the bilocal operator is obtained between the
vacuum and $B(v)$-state as:
\begin{eqnarray}\label{eq2160}
\widetilde{\Pi}_{\mu\nu}^{V-A }(p,\widetilde{q})&=&  i\int d^4x~
e^{ip.x}\times {\left\{\gamma_{\mu}\gamma_{5}~ iS_d(x)~ \gamma_{\nu}
(1- \gamma_{5})\right\}}_{\alpha\beta}\langle0|\bar{u}_{\alpha}(x)
h_{v\beta}(0) |B(v)\rangle, \nonumber\\
\widetilde{\Pi}_{\mu\nu}^{T}(p,\widetilde{q})&=&  i\int d^4x~
e^{ip.x}\times {\left\{\gamma_{\mu}\gamma_{5}~ iS_d(x)~
\sigma_{\nu\eta}q^\eta
(1+\gamma_5)\right\}}_{\alpha\beta}\langle0|\bar{u}_{\alpha}(x)
h_{v\beta}(0) |B(v)\rangle.
\end{eqnarray}

The full-quark propagator, $S_d(x)$ of a massless quark in the
external gluon field in the Fock-Schwinger gauge is as follows:
\begin{eqnarray}\label{eq217}
S_d(x) &=& i \int \frac{d^{4}k}{(2\pi)^{4}}
e^{-ik.x}\left\{\frac{\not\! k}{k^{2}}-\int^{1}_{0}dv~ G_{\mu\nu}(v
x)\left[\frac{1}{2k^{4}}\not\! k \sigma^{\mu\nu}-\frac{1}{k^2}
vx^{\mu}\gamma^{\nu}\right] \right\}.
\end{eqnarray}
In addition, the DA's of the $B$-meson are as \cite{Wang}:
\begin{eqnarray}\label{eq218}
\langle 0|\bar {u}{_{\alpha}(x){h_{v\beta}(0)}}|B(v)\rangle &=&
-\frac{if_{B}m_{B}}{4} \int^{\infty}_{0} d\omega~ e^{-i\omega v.x}
\left\{ (1+\not\! v)\varphi_{_+}-\frac{\not\! x\gamma_{5}}{2v
x}(\varphi_{_+}-\varphi_{_-})\right\}_{\beta\alpha},
\nonumber\\
\langle 0|\bar{u}_\alpha(x) G_{\lambda\rho}(ux)
h_{v\beta}(0)|B(v)\rangle &=& \frac{f_Bm_B}{4}\int_0^\infty d\omega
\int_0^\infty d\xi  e^{-i(\omega+u\xi) v .x} \left\{
\left[(v_\lambda\gamma_\rho-v_\rho\gamma_\lambda)
\Big(\Psi_A-\Psi_V\Big)
\right. \right. \nonumber\\
&&\left. \left. -i\sigma_{\lambda\rho}\Psi_V-\frac{x_\lambda
v_\rho-x_\rho v_\lambda}{v x}X_A+\frac{x_\lambda \gamma_\rho-x_\rho
\gamma_\lambda}{v x}Y_A\right](1 +
\not\!v)\gamma_5\right\}_{\beta\alpha},\nonumber\\
\end{eqnarray}
where \cite{Mannel}:
\begin{eqnarray}\label{eq219}
\Psi_{A}(\omega , \xi) &=&   \Psi_{V}(\omega ,
\xi)=\frac{\lambda_{E}^{2}}{6\omega_{0}^4}\,\xi^{2}\,e^{-\frac{\omega+\xi}{\omega_{0}}},
\nonumber\\
X_{A}(\omega , \xi)   &=&
\frac{\lambda_{E}^{2}}{6\omega_{0}^4}\,\xi(2\omega-\xi)\,e^{-\frac{\omega+\xi}{\omega_{0}}},
\nonumber\\
Y_{A}(\omega, \xi)   &=&
-\frac{\lambda_{E}^{2}}{24\omega_{0}^4}\,\xi(7\omega_{0}-13\omega+3\xi)\,e^{-\frac{\omega+\xi}{\omega_{0}}}.
\end{eqnarray}
Our knowledge of the behaviors of $\varphi_{\pm}(\omega)$ at small
$\omega$ is still rather limited due to the poor understanding of
non-perturbative QCD dynamics. To achieve a better understanding of
the model dependence of $\varphi_{\pm}(\omega)$ in the sum rule
analysis, we consider the following four different parameterizations
for the shapes of the $B$-meson DA $\varphi_{+}$
\cite{Grozin,Braun,Faziophi,Wangphi}:
\begin{eqnarray}\label{eq2020}
\varphi_{+,\rm I}(\omega) &=& \frac{\omega}{\omega_0^2} \,
e^{-\omega/\omega_0} \,,
\nonumber \\
\varphi_{+,\rm II}(\omega) &=& \frac{1}{4 \pi \,\omega_0} \, {k
\over k^2+1} \, \left[ {1 \over k^2+1} - \frac{2 (\sigma_B
-1)}{\pi^2}  \, \ln k \right ] \,, \hspace{1 cm} k= \frac{\omega}{1
\,\, \rm GeV} \,, \,
\nonumber \\
\varphi_{+, \rm III}(\omega) &=& \frac{2 \omega^2}{\omega_0
\omega_1^2} \, e^{-(\omega/\omega_1)^2} \,, \hspace{1 cm} \omega_1=
{2 \, \omega_0 \over \sqrt{\pi}} \,,
\nonumber \\
\varphi_{+,\rm IV}(\omega) &=& \frac{\omega }{\omega_0 \omega_2} \,
{\omega_2 - \omega \over \sqrt{\omega(2 \omega_2-\omega)} } \,\,
\theta(\omega_2-\omega)\,, \hspace{1 cm} \omega_2= {4 \, \omega_0
\over 4- \pi} \,.
\end{eqnarray}
The determination of coefficient $\omega_0$, which constitutes the
most important theory uncertainty in the $B$-meson LCSR approach,
will be discussed for each of the four models in the next section.

The corresponding expression of $\varphi_{-}(\omega)$  for each
model is determined by:
\begin{eqnarray}\label{eq2021}
\varphi_{-}(\omega) = \int_0^1 \, { d \xi \over \xi} \,
\varphi_{+}\left (\omega/ \xi \right ).
\end{eqnarray}
These parameterizations can provide a reasonable description of
$\varphi_{\pm}(\omega)$ at small $\omega$ due to the radiative tail
developed from QCD corrections.

Inserting  the full propagator  and $B$-meson DA's presented in Eqs.
(\ref{eq217}) and (\ref{eq218}), respectively, in the correlation
functions (Eq. (\ref{eq2160})), traces and then integrals should be
calculated. To estimate these calculations, we have used $x_\mu
\rightarrow i \frac{\partial}{\partial{k_\mu}}$. In addition to
this, for terms containing a factor of $vx$ in the denominator, we
have used the following trick: in order not to have any singularity
at $v.x = 0$, the integral of these wave functions in the absence of
the exponential should cancel. Hence, for these terms only, one can
write:
\begin{eqnarray}\label{eq220}
e^{i\alpha v. x}\rightarrow e^{i\alpha v. x}-1=iv. x
\int_{0}^{\alpha} dk~ e^{ik v. x},
\end{eqnarray}
and the rest of the calculation is similar to the presented one.
Note that the subtracted $1$ does not contribute.

After completing the integrals and matching them with the hadronic
representation below the continuum threshold $s _{0}$, through the
dispersion relation and applying Borel transform with respect to the
variable $p^2$ as:
\begin{eqnarray}\label{eq221}
B_{p^{2}}(M^{2})(\frac{1}{p^{2}-m^{2}})^{n}&=&\frac{(-1)^{n}}{\Gamma(n)}\frac{e^{-\frac{m^{2}}{M^{2}}}}{(M^{2})^{n}},
\end{eqnarray}
in order to suppress the contributions of the higher states, the
form factors are obtained via the LCSR. For instance, the form
factor $V_{1}$ is presented here:
\begin{eqnarray}\label{eq222}
V_{1}(q^{2})&=&-\frac{f_{B}m_{B}}{f_{a}m_{a_1}(m_{B}-m_{a_1})}e^\frac{m_{a_1}^2}{M^2}\left\{\int_{0}^{\sigma_{0}}d\sigma\left[\frac{M^{2}}{2} \varphi_{_+}(\omega^{'}) \frac{d}{d\sigma}e^{-\frac{s}{M^{2}}}\right]\right.
+\hat{\mathcal L}\left[(1-u)\Psi_{_V}~\frac{d}{d\sigma}e^{-\frac{s}{M^{2}}} \right.\nonumber\\
&-&\frac{2u-1}{2M^{2}}(\Psi_{_A}-\Psi_{_V})~\frac{d}{d\sigma}e^{-\frac{s}{M^{2}}}
+\frac{m_{B}^{2}(1-2u)}{\bar{\sigma}^2{M^2}}(3\widetilde{X}_{_A}-\widetilde{Y}_{_A})~e^{-\frac{s}{M^{2}}}+\frac{1}{\bar{\sigma}^3M^2} \widetilde{X}_{_A}~ e^{-\frac{s}{M^2}}    \nonumber\\
&\times&\left.\left. \left(m_{B}^2
(1+3\sigma)-2\widetilde{m}{_{B}^2}+\frac{\mathcal
M^{4}}{2M^2}\right) \right]\right\},
\end{eqnarray}
The explicit expressions for the other form factors are presented in
Appendix.

Finally, with a little bit of change in the previous steps, such as
the change in the quark spectator  ($u \to s$),  we can easily find
similar results for the form factors of the $B_s \to K_{1A}$, and
$B_s \to K_{1B}$ decays.

The form factors of $B_{s} \to K_1(1270) \ell^+ \ell^- $ transitions
with the mixing angle $\theta_K$ are defined as \cite{Hua}
\begin{eqnarray}\label{eq234}
f^{K_{1}(1270)}&=&C_1\,\sin\theta_K\,f^{K_{1A}}+C_2\,\cos\theta_K\,f^{K_{1B}},
\end{eqnarray}
where $f^{K_{1}(1270)}, f^{K_{1A}}$ and $f^{K_{1B}}$ stand for the
form factors $A, V_i (i=0,1,2), T_j (j=1,2,3)$ of $B_{s} \to
K_1(1270), B_s \to K_{1A}$ and $B_s \to K_{1B}$ decays,
respectively. The coefficients $C_1$ and $C_2$ related to each form
factor of $B_s \to K_1(1270)$ decay are given in Table \ref{T0}.
\begin{table}[th]
\caption{The coefficients  $C_1$ and $C_2$ for each form factor of
$B_s \to K_1(1270)$ .} \label{T0}
\begin{ruledtabular}
\begin{tabular}{ccc}
\rm{Form factors} &$C_1$&$C_2$ \\
\hline
$A, V_1, V_2$ & $\frac{m_{B_s}-m_{K_1}}{m_{B_s}-m_{K_{1A}}}$ & $\frac{m_{B_s}-m_{K_1}}{m_{B_s}-m_{K_{1B}}}$  \\
$V_0$         & $\frac{m_{K_{1A}}}{m_{K_{1}}}$ & $\frac{m_{K_{1B}}}{m_{K_{1}}}$  \\
$T_1, T_3$    & $1$ & $1$  \\
$T_2$         &
$\frac{{m_{B_{s}}^2}-m_{K_{1A}}^2}{m_{B_{s}}^2-m_{K_{1}}^2}$ &
$\frac{m_{B_{s}}^2-m_{K_{1B}}^2}{m_{B_{s}}^2-m_{K_{1}}^2}$
\end{tabular}
\end{ruledtabular}
\end{table}

\section{ Numerical Analysis }\label{sec3}
In this section, our numerical analysis of the form factors $A$,
$V_{i}$ and $T_{j}$ are presented for the $B_{(s)}\to a_1(K_1)
\ell^+ \ell^-$ decays. The values are chosen for masses in
$\mbox{GeV}$  as $m_{B}=(5.27\pm 0.01)$, $m_{a_1}=(1.23\pm 0.04)$,
$m_{K_1}=(1.27\pm 0.01)$,  $m_{\mu}=0.11$ and $m_{\tau}=1.77$
\cite{pdg}, $m_{K_{1A}}=(1.31\pm 0.06)$,  $m_{K_{1B}}=(1.34\pm
0.08)$ \cite{Yang}. The leptonic decay constants are taken as:
$f_{a_1}=(0.24\pm 0.01)\, \mbox{GeV}$, $f_{K_{1A}}=(0.25\pm 0.01)\,
\mbox{GeV}$, $f_{K_{1B}}=(0.19\pm 0.01)\, \mbox{GeV}$ \cite{Yang},
$f_{B}=(0.18\pm 0.02)\, \mbox{GeV}$ \cite{Wan}, and
$f_{B_s}=(0.23\pm 0.03)\, \mbox{GeV}$ \cite{Li2}.  Moreover, $s
_{0}=(2.55\pm0.15) \, \mbox{GeV}^2$ is used for the continuum
threshold \cite{Yang} . The values of the parameters $\lambda
_{E}^{2}$ and $\sigma_B$ of the $B$-meson DA's are chosen as
$\lambda _{E}^{2}=(0.11\pm0.06) \,\mbox{GeV}^2$ \cite{Grozin} and
$\sigma_B= 1.4\pm 0.4$ \cite{Braun}.  The Borel parameter in this
article is taken as $1.5 ~\rm GeV^2 \leq M^2 \leq 4~\rm GeV^2$. In
this region, the values of the  form factors $A$, $V_{i}$ and
$T_{j}$ are stable enough. The uncertainties which originated from
the Borel parameter $M^2$ in this  interval, are about $1 \%$.

\begin{figure}[th]
\includegraphics[width=5cm,height=6cm]{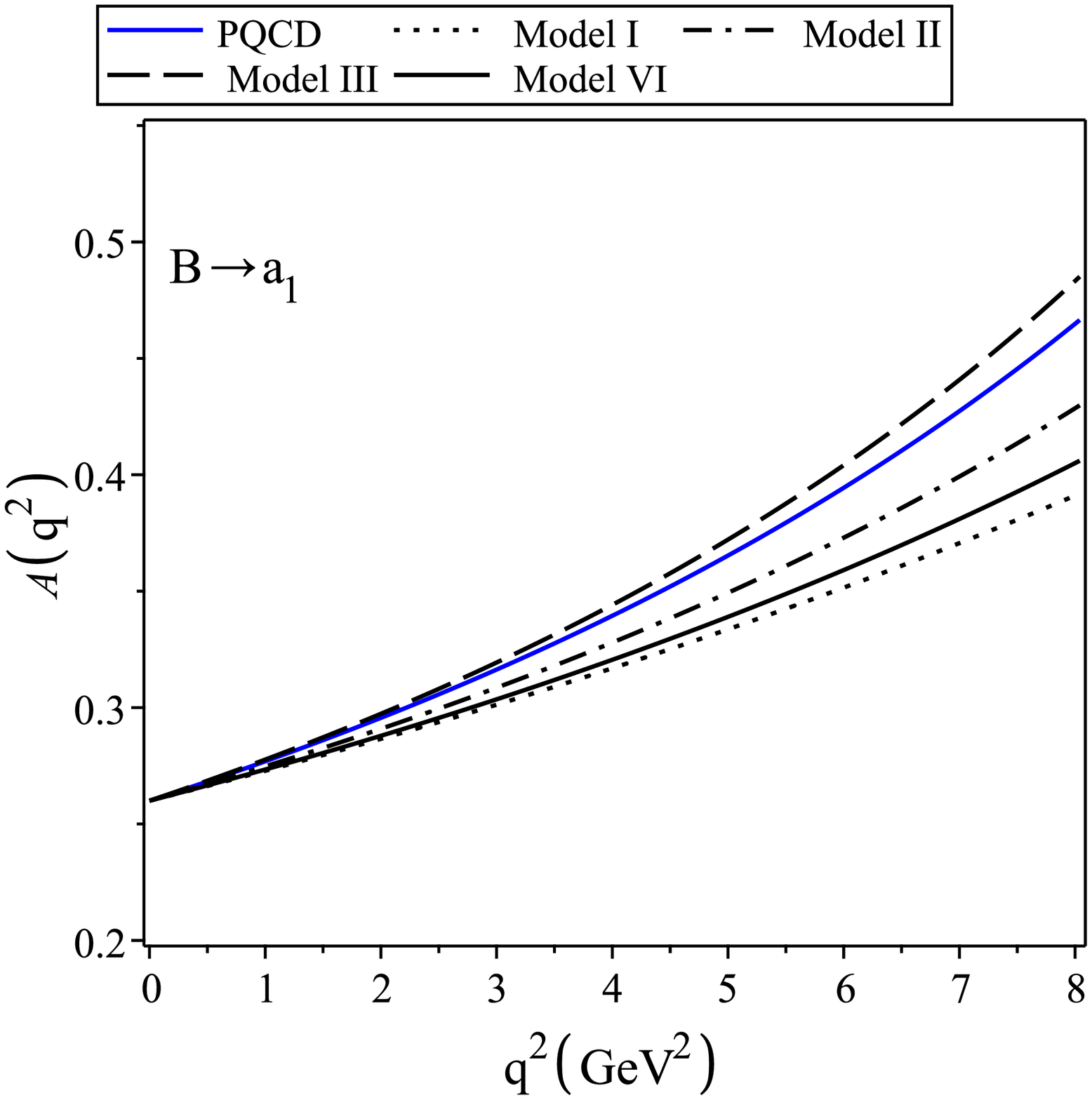}
\includegraphics[width=5cm,height=6cm]{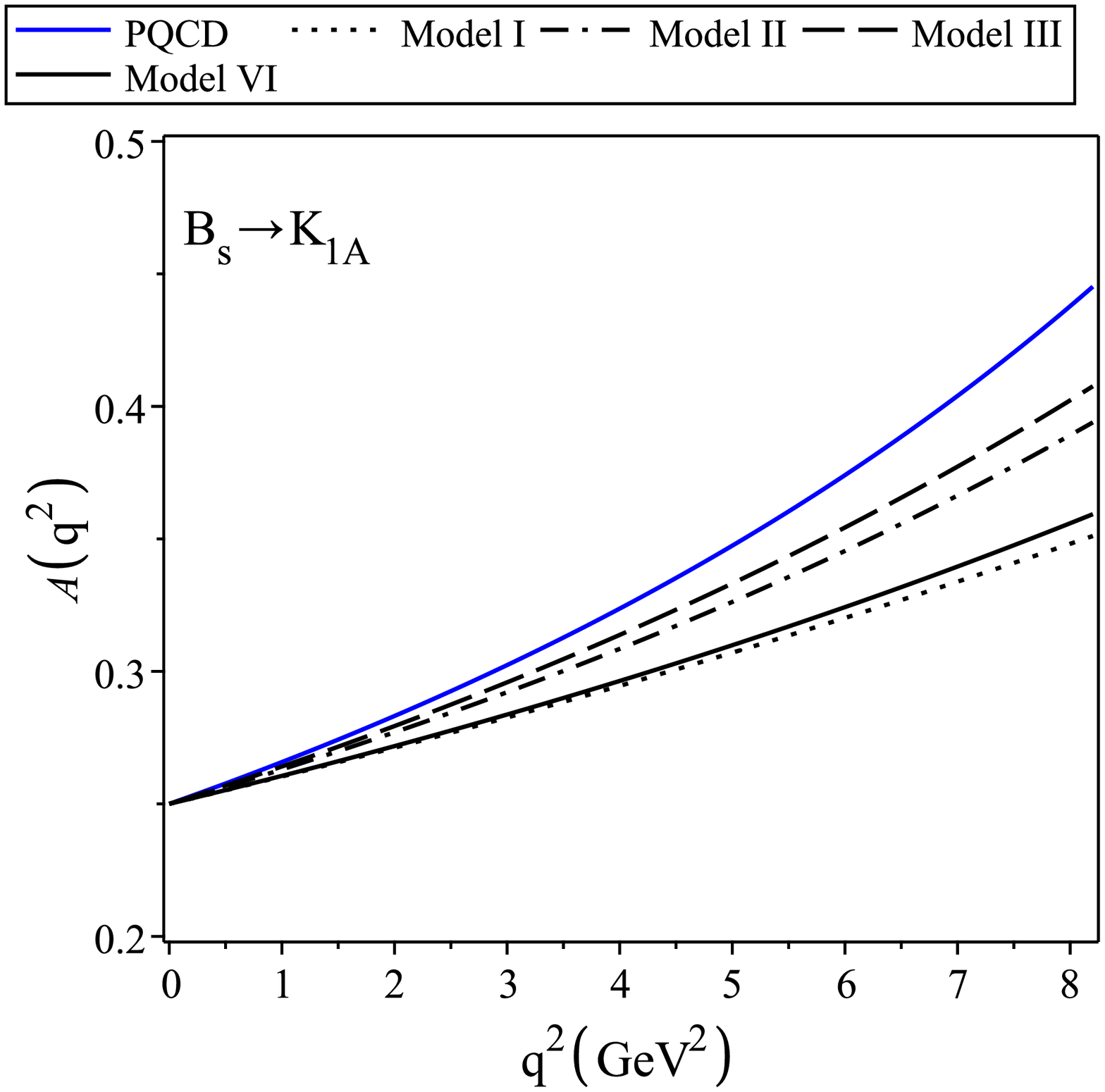}
\includegraphics[width=5cm,height=6cm]{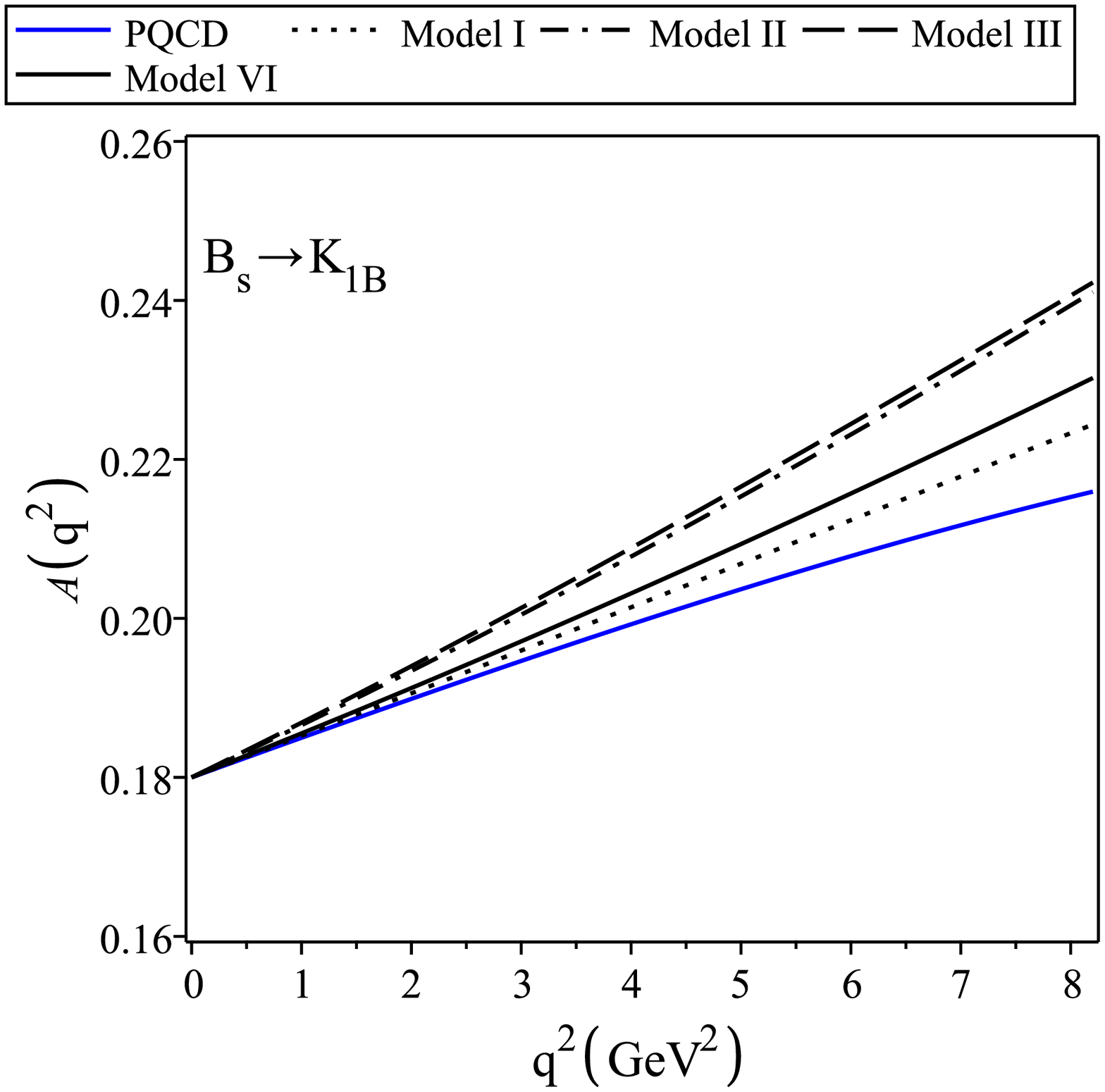}
\caption{Dotted, dot-dashed, dashed, solid (black) curves show the
form factors $A$, calculated with the $B$-meson DA's, whose values
at $q^2=0$ have been fixed to the prediction from the PQCD (blue).}
\label{F41}
\end{figure}

Having all these input values and parameters at hand, we proceed to
carry out numerical calculations. As can be seen in Eq.
(\ref{eq2020}), the $B$-meson DA's, $\varphi_{\pm}$ in the four
cases are related to the parameter $\omega_0$ whose value is depend
on the model. In order to determine the parameter $\omega_0$ for
$B\to a_1 \ell^+ \ell^-$ decay, we match the values of the form
factor $A^{B\to a_1}$ in $q^2=0$,  estimated with the four models of
the $B$-meson DA's $\varphi_{\pm}$, with $A^{B\to a_1}(0) = 0.26\pm
0.09$ computed from the PQCD as a different method \cite{Li2}, and
derive the values of the coefficient $\omega_0$ for each model.
Also, taking $A^{B_s\to K_{1A}}(0) = 0.25\pm {0.10}$ and $A^{B_s\to
K_{1B}}(0) = 0.18\pm {0.08}$ evaluated via the PQCD \cite{Li2}, and
performing the same procedure as $B\to a_1$ decay for $B_{s}\to
K_{1A}$ and $B_{s}\to K_{1B}$ transitions, the values of the
parameter $\omega_0$ are calculated for these decays. The values of
the parameter $\omega_0$ for three aforementioned decays are given
in Table \ref{T41}.
\begin{table}[th]
\caption{The values of  $\omega_0$ for each model in MeV .}
\label{T41}
\begin{ruledtabular}
\begin{tabular}{ccccc}
\rm{Model}&\rm{I} &\rm{II} &\rm{III} &\rm{IV}  \\
\hline $\omega_0 \,({\rm for}~  B \to a_1)$  & $235^{+25}_{-19} $ &
$246^{+28}_{-21} $ &
$259^{+29}_{-22}$ &  $217^{+20}_{-16}$\\
$\omega_0\,({\rm for}~  B_{s}\to K_{1A})$ & $254^{+27}_{-20} $ &
$267^{+29}_{-22} $ & $281^{+33}_{-25}$ & $234^{+23}_{-17}$\\
$\omega_0\,({\rm for}~  B_{s}\to K_{1B})$ & $282^{+33}_{-22} $ &
$298^{+34}_{-26} $ & $313^{+37}_{-28}$ & $259^{+23}_{-20}$
\end{tabular}
\end{ruledtabular}
\end{table}
Fig. \ref{F41} shows the form factors $A^{B\to a_1}$,  $A^{B_{s}\to
K_{1A}}$ and $A^{B_{s}\to K_{1B}}$ with the four models of the
$B$-meson DA's $\varphi_{\pm}$ whose values at zero momentum
transfer have been fixed to the predictions from the PQCD. In this
figure, blue lines show the form factors predicted by the PQCD
method.

Now, by inserting the values of the masses, leptonic decay
constants, continuum threshold, Borel mass, the parameters of the
$B$-meson DA's such as $\omega_0$ and other quantities that appear
in the form factors, we can calculate the form factors of
$B_{(s)}\to a_1(K_{1A}, K_{1B})$ decays at zero momentum transfer.
Taking into account all the uncertainties, the numerical values of
the form factors $A$, $V_{i}$ and $T_{j}$ for aforementioned decays
in $q^2=0$ are presented in Table \ref{T42} for the four models of
the $B$-meson DA's, $\varphi_{\pm}$.  The main uncertainty comes
from $\omega_{0}$, the decay constant $f_{a_1}$  $(f_{K_{1A}},
f_{K_{1B}}$), and $B$-meson mass.
\begin{table}[th]
\caption{The $B_{(s)} \to a_1(K_{1A}, K_{1B})$   form factors at
zero momentum transfer in the four models of $B$-meson DA's,
$\varphi_{\pm}$.} \label{T42}
\begin{ruledtabular}
\begin{tabular}{cccccccc}
&$\mbox{Model}$&${A^{B\to a_1}}$&${V}^{B\to a_1}_{1}$&$V^{B\to a_1}_{2}$&$V^{B\to a_1}_{0}$&$T^{B\to a_1}_{1}= T^{B\to a_1}_{2}$&$T^{B\to a_1}_{3}$ \\
\hline
&$\mbox {I}$ & ${0.26}{\pm 0.09} $ & ${0.42}{\pm 0.13} $ &${0.22}{\pm 0.07} $ &${0.11}{\pm 0.03} $ &${0.25}{\pm 0.08} $&${0.22}{\pm 0.06} $\\
&$\mbox {II} $ & ${0.26}{\pm 0.09} $ & ${0.51}{\pm 0.16} $ &${0.25}{\pm 0.08} $&${0.13}{\pm 0.04} $&${0.30}{\pm 0.09} $&${0.25}{\pm 0.07} $\\
&$\mbox {III}$ & ${0.26}\pm{0.09} $&${0.54}\pm{0.17} $&$ {0.28}\pm{0.09}$&${0.27}\pm{0.08} $& ${0.34}\pm{0.11}$&${0.29}\pm{0.09}$\\
&$\mbox {VI}$ & ${0.26}\pm{0.09} $&${0.40}\pm{0.13} $&$ {0.20}\pm{0.06}$&${0.11}\pm{0.03} $& ${0.24}\pm{0.07}$&${0.21}\pm{0.06}$\\
\hline
&$\mbox{Model}$&${A^{B_{s}\to K_{1A}}}$&${V}^{B_{s}\to K_{1A}}_{1}$&$V^{B_{s}\to K_{1A}}_{2}$&$V^{B_{s}\to K_{1A}}_{0}$&$T^{B_{s}\to K_{1A}}_{1}= T^{B_{s}\to K_{1A}}_{2}$&$T^{B_{s}\to K_{1A}}_{3}$ \\
\hline
&$\mbox {I}$ & ${0.25}{\pm 0.10} $ & ${0.35}{\pm 0.12} $ &${0.15}{\pm 0.05} $ &${0.12}{\pm 0.04} $ &${0.27}{\pm 0.09} $&${0.24}{\pm 0.08} $\\
&$\mbox {II} $ & ${0.25}{\pm 0.10} $ & ${0.37}{\pm 0.13} $ &${0.18}{\pm 0.05} $&${0.11}{\pm 0.03} $&${0.32}{\pm 0.11} $&${0.27}{\pm 0.09} $\\
&$\mbox {III}$ & ${0.25}\pm{0.10} $&${0.45}\pm{0.15} $&$ {0.22}\pm{0.07}$&${0.12}\pm{0.04} $& ${0.39}\pm{0.13}$&${0.33}\pm{0.12}$\\
&$\mbox {VI}$ & ${0.25}\pm{0.10} $&${0.33}\pm{0.11} $&$ {0.13}\pm{0.04}$&${0.11}\pm{0.03} $& ${0.25}\pm{0.09}$&${0.21}\pm{0.07}$\\
\hline
&$\mbox{Model}$&${A^{B_s\to K_{1B}}}$&${V}^{B_s\to K_{1B}}_{1}$&$V^{B_s\to K_{1B}}_{2}$&$V^{B_s\to K_{1B}}_{0}$&$T^{B_s\to K_{1B}}_{1}= T^{B_s\to K_{1B}}_{2}$&$T^{B_s\to K_{1B}}_{3}$ \\
\hline
&$\mbox{I}$&$0.18\pm0.08 $ & $0.28\pm 0.10 $ &$0.12\pm 0.04 $ & $0.07\pm 0.02$ & $0.22\pm 0.08 $ &$0.19\pm 0.07 $\\
&$\mbox{II}$&$0.18\pm0.08 $ & $0.33\pm 0.12 $ &$0.15\pm 0.05 $ & $0.10\pm 0.03$ & $0.28\pm 0.10 $ &$0.25\pm 0.09 $\\
&$\mbox{III}$&$0.18\pm0.08 $ & $0.38\pm 0.14 $ &$0.20\pm 0.07 $ & $0.12\pm 0.03$ & $0.33\pm 0.11 $ &$0.30\pm 0.11 $\\
&$\mbox{VI}$&$0.18\pm0.08 $ & $0.27\pm 0.08 $ &$0.11\pm 0.03 $ & $0.09\pm 0.02$ & $0.21\pm 0.07 $ &$0.18\pm 0.05 $\\
\end{tabular}
\end{ruledtabular}
\end{table}

So far, several authors have calculated the form factors of the
$B_{(s)}\to a_{1}(K_{1A}, K_{1B}) \ell^{+} \ell^{-}$ decays via
differen frameworks. To compare the results, we should rescale them
according to the form factor definitions in Eq. (\ref{eq211}). Table
\ref{T43} shows the values of the rescaled  form factors at $q^2=0$
from different approaches.
\begin{table}[th] \caption{Transition form factors of
the $B_{(s)}\to a_{1}(K_{1A}, K_{1B}) \ell^+ \ell^-$ at $q^2=0$ in
various methods. } \label{T43}
\begin{ruledtabular}
\begin{tabular}{ccccccccc}
$\mbox{Approaches}$&$A^{B\to a_1}$&$V^{B\to a_1}_{1}$&$V^{B\to a_1}_{2}$&$V^{B\to a_1}_{0}$&$T^{B\to a_1}_{1}= T^{B\to a_1}_{2}$&$T^{B\to a_1}_{3}$ \\
\hline
PQCD \cite{Li2}  &$ 0.26\pm0.09$&$ 0.43\pm0.16$&  $ 0.13\pm0.04$ & $0.34\pm0.16 $&$ 0.34\pm0.13$&$0.30\pm0.17$  \\
LCSR \cite{mo} &$0.42\pm0.16$&$0.68\pm0.13$&$0.31\pm0.16$&$0.30\pm0.18$&$0.44\pm0.28$&$0.41\pm0.18$  \\
3PSR \cite{kh} &$0.51$&$0.52$&  $0.25$ & $0.76$&$0.37$&$0.41$  \\
\hline
&${A^{B_s\to K_{1A}}}$&${V}^{B_s\to K_{1A}}_{1}$&$V^{B_s\to K_{1A}}_{2}$&$V^{B_s\to K_{1A}}_{0}$&$T^{B_s\to K_{1A}}_{1}= T^{B_s\to K_{1A}}_{2}$&$T^{B_s\to K_{1A}}_{3}$ \\
\hline
PQCD \cite{Li2}  &$ 0.25 \pm 0.10 $ & $ 0.43 \pm 0.19 $ & $ 0.11 \pm 0.05 $ & $ 0.36 \pm 0.18 $ & $ 0.34 \pm 0.15 $ & $ 0.30 \pm 0.13$  \\
\hline
&${A^{B_s\to K_{1B}}}$&${V}^{B_s\to K_{1B}}_{1}$&$V^{B_s\to K_{1B}}_{2}$&$V^{B_s\to K_{1B}}_{0}$&$T^{B_s\to K_{1B}}_{1}= T^{B_s\to K_{1B}}_{2}$&$T^{B_s\to K_{1B}}_{3}$ \\
\hline
PQCD \cite{Li2}  &$ 0.18 \pm 0.08$ & $ 0.33\pm 0.14$ & $ 0.03\pm 0.03$ & $ 0.42\pm 0.16$ & $ 0.26\pm 0.11$ & $ 0.17\pm 0.08$  \\
\end{tabular}
\end{ruledtabular}
\end{table}
Considering the uncertainties, our results for the form factors of
these  decays  are in a good agreement with those of the PQCD in
most cases (except $V_2$ and $V_0$). However, there is not good
agreement between our results with the LCSR with $a_1$-meson DA's
\cite{mo}.

The LCSR calculations for the form factors are truncated at about $0
\leq q^2 \leq 8 {\rm GeV}^2$. To extend the $q^2$ dependence of the
form factors to the full physical region, where the LCSR results are
not valid, we find that the sum rules predictions for the form
factors are well fitted to the following function:
\begin{eqnarray}\label{eq31}
F_{i}(q^{2})=\frac{F_i(0)}{1-\alpha(q^2/m_{B_{(s)}}^2)+\beta
{(q^2/m_{B_{(s)}}^2)}^2},
\end{eqnarray}
where  $F_{i}(0)$,  $\alpha$ and $\beta$ are the constant fitted
parameters. The values of the parameters $[\alpha, \beta]$ are
presented in Tables \ref{T44}, \ref{T45} and  \ref{T46} for $B \to
a_1$, $B_{s}\to K_{1A}$ and $B_{s}\to K_{1B}$, respectively.  The
values of parameter $F_{i}(0)$ expressed the form factor results at
$q^2=0$ were listed in Table \ref{T42}, before.
\begin{table}[th]
\caption{The parameters $[\alpha, \beta]$ obtained for the form
factors of the $B\to a_1$ transition in the four models.}
\label{T44}
\begin{ruledtabular}
\begin{tabular}{ccccccccc}
$\mbox{Model}$& ${A}(q^2)$&${V}_{1}(q^2)$&${V}_{2}(q^2)$&${V}_{0}(q^2)$&$T_{1}(q^2)$&$T_{2}(q^2)$&$T_{3}(q^2)$ \\
\hline I&
$[1.33,0.57]$&$[0.87, 0.38]$&$[1.01, 0.53]$&$[1.04, 0.29]$&$[1.11, 0.21]$&$[1.25, 0.48]$&$[1.13, 1.10]$\\
II&$[1.51, 0.50]$&$[1.15, 0.36]$&$[1.38, 0.54]$&$[1.28, 0.26]$&$[1.29, 0.47]$&$[1.23, 0.29]$&$[1.29, 0.94]$\\
III&$[1.70, 0.64]$&$[1.20, 0.46]$&$[1.28, 0.73]$&$[1.15, 0.31]$&$[1.36, 0.52]$&$[1.30, 0.35]$&$[1.62, 0.87]$\\
IV&
$[1.34, 0.46]$&$[0.88, 0.30]$&$[1.03, 0.42]$&$[1.08, 0.22]$&$[1.41, 0.17]$&$[1.31, 0.38]$&$[1.14, 0.88]$
\end{tabular}
\end{ruledtabular}
\end{table}
\begin{table}[th]
\caption{The same as Table \ref{T44} but for $B_{s}\to K_{1A}$
transition.} \label{T45}
\begin{ruledtabular}
\begin{tabular}{ccccccccc}
$\mbox{Model}$& ${A}(q^2)$&${V}_{1}(q^2)$&${V}_{2}(q^2)$&${V}_{0}(q^2)$&$T_{1}(q^2)$&$T_{2}(q^2)$&$T_{3}(q^2)$ \\
\hline I&
$[1.16, 0.53]$&$[0.98, 0.24]$&$[1.14, 0.39]$&$[0.96, 0.34]$&$[1.30, 0.17]$&$[1.09, 0.46]$&$[0.95, 1.17]$\\
II&
$[1.44, 0.56]$&$[1.31, 0.32]$&$[1.20, 0.49]$&$[1.25, 0.34]$&$[1.34, 0.28]$&$[1.33, 0.35]$&$[1.08, 0.91] $\\
III&
$[1.56, 0.72]$&$[1.52, 0.31]$&$[1.37, 0.76]$&$[1.32, 0.43]$&$[1.46, 0.50]$&$[1.38, 0.39]$&$[1.12, 0.98] $\\
VI&
$[1.18, 0.42]$&$[1.01, 0.45]$&$[1.15, 0.31]$&$[0.99, 0.27]$&$[1.33, 0.13]$&$[1.11, 0.36]$&$[0.96, 1.03]$
\end{tabular}
\end{ruledtabular}
\end{table}

\begin{table}[th]
\caption{The same as Table \ref{T44} but for $B_{s}\to K_{1B}$
transition. } \label{T46}
\begin{ruledtabular}
\begin{tabular}{ccccccccc}
$\mbox{Model}$& ${A}(q^2)$&${V}_{1}(q^2)$&${V}_{2}(q^2)$&${V}_{0}(q^2)$&$T_{1}(q^2)$&$T_{2}(q^2)$&$T_{3}(q^2)$ \\
\hline I&
$[0.83, 0.48]$&$[0.62, 0.29]$&$[0.75, 0.49]$&$[0.77, 0.37]$&$[0.96, 0.31]$&$[0.78, 0.37]$&$[0.63, 1.47]$\\
II&
$[1.03, 0.50]$&$[0.81, 0.35]$&$[0.81, 0.41]$&$[0.93, 0.40]$&$[0.99, 0.28]$&$[0.99, 0.28]$&$[0.71, 1.14]$\\
III &
$[1.08, 0.63]$&$[0.90, 0.28]$&$[0.80, 0.77] $&$[0.96, 0.42]$&$[1.01, 0.56]$&$[0.91, 0.24]$&$[0.38, 1.04] $\\
VI&
$[0.87, 0.37]$&$[0.66, 0.22]$&$[0.81, 0.39]$&$[0.82, 0.29]$&$[0.98, 0.23]$&$[0.84, 0.28]$&$[0.66, 1.18]$
\end{tabular}
\end{ruledtabular}
\end{table}

\begin{figure}[th]
\includegraphics[width=6cm,height=6cm]{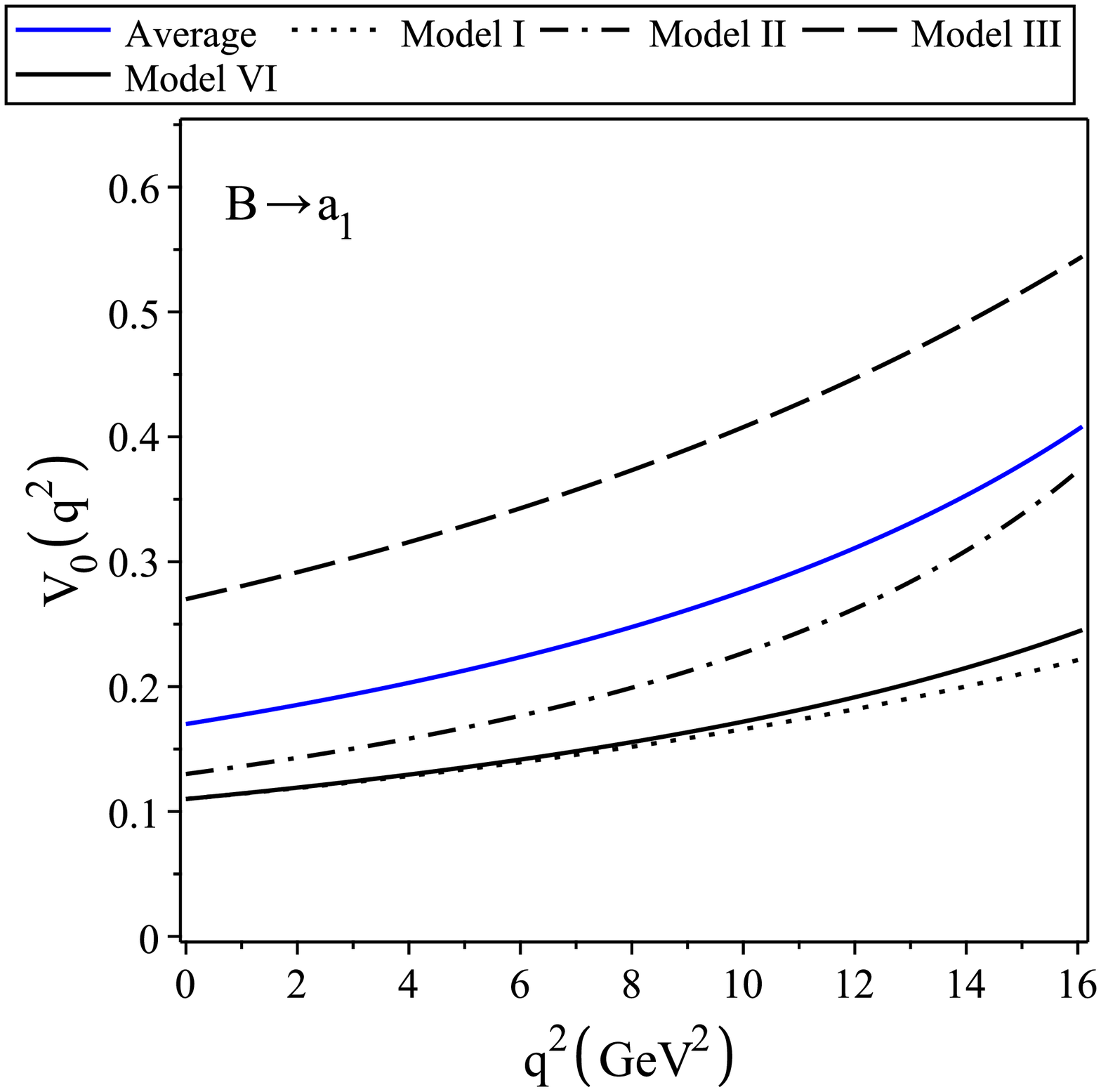}
\includegraphics[width=6cm,height=6cm]{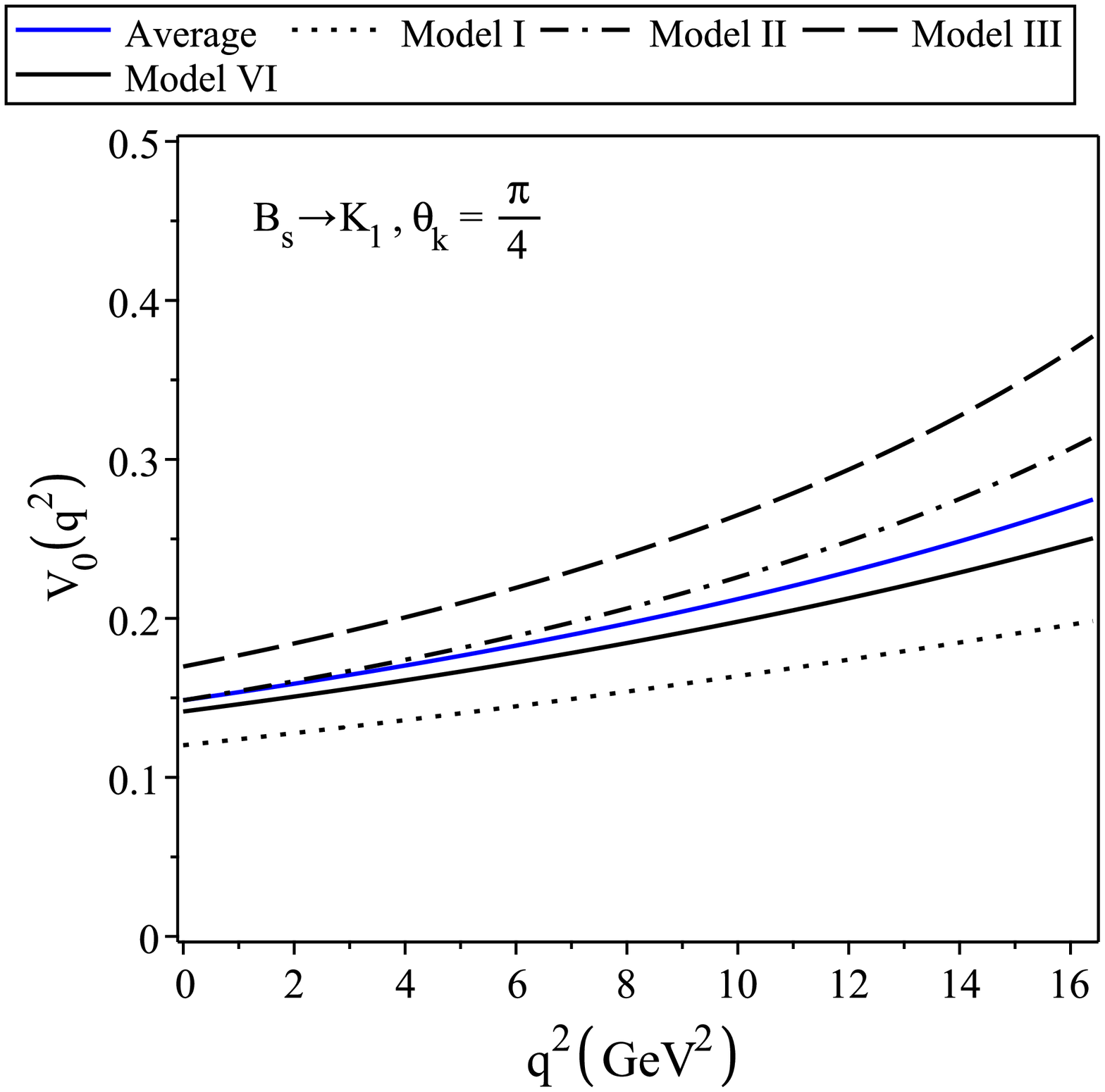}
\includegraphics[width=6cm,height=6cm]{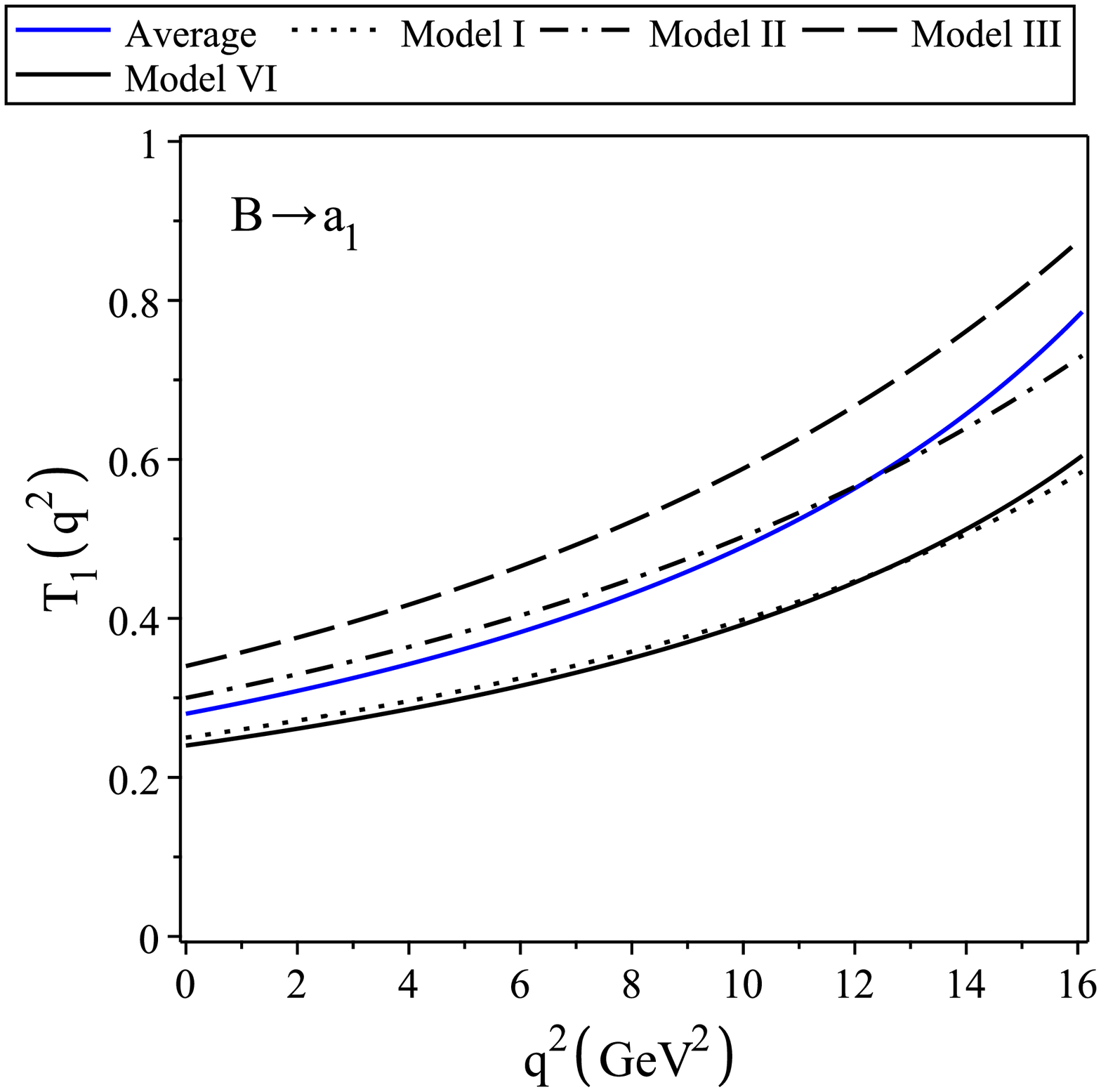}
\includegraphics[width=6cm,height=6cm]{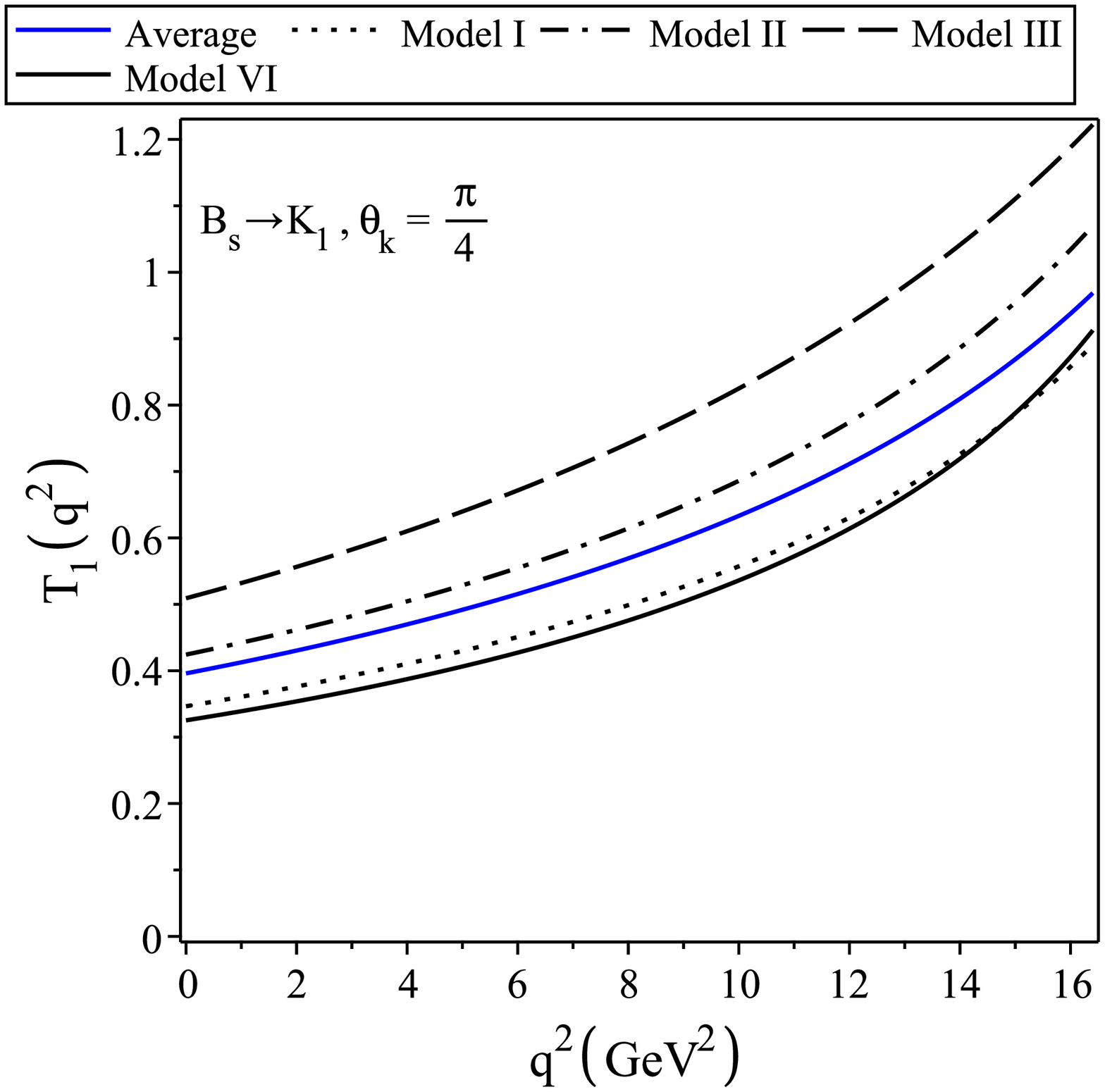}
\caption{The form factors ${V}_{0}$ and ${T}_{1}$ of $B_{(s)}\to
a_1(K_1)$ on $q^2$ with the four models (black color). Blue lines
show the average form factors. } \label{F42}
\end{figure}

\begin{figure}[th]
\includegraphics[width=6cm,height=6cm]{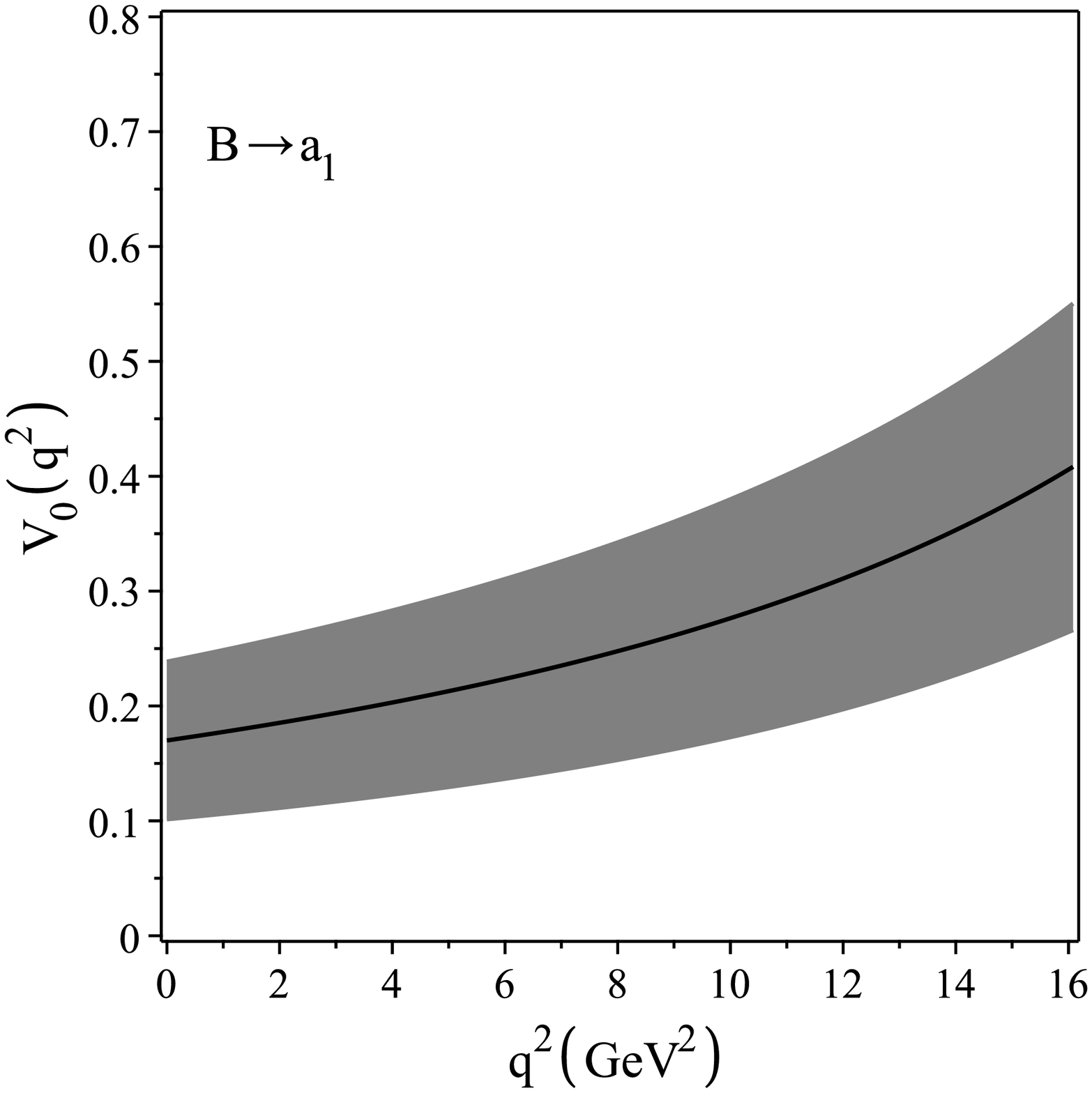}
\includegraphics[width=6cm,height=6cm]{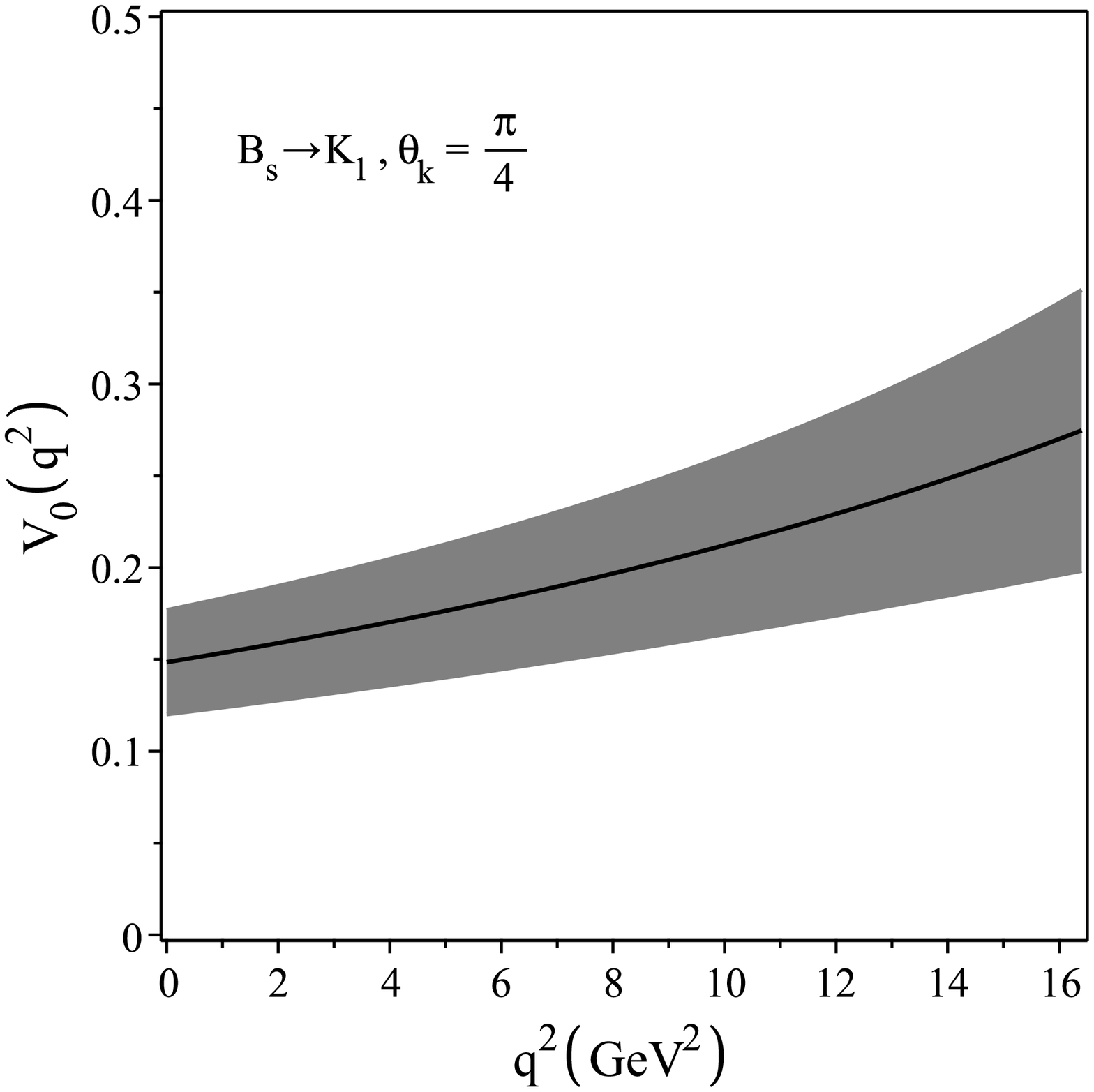}
\includegraphics[width=6cm,height=6cm]{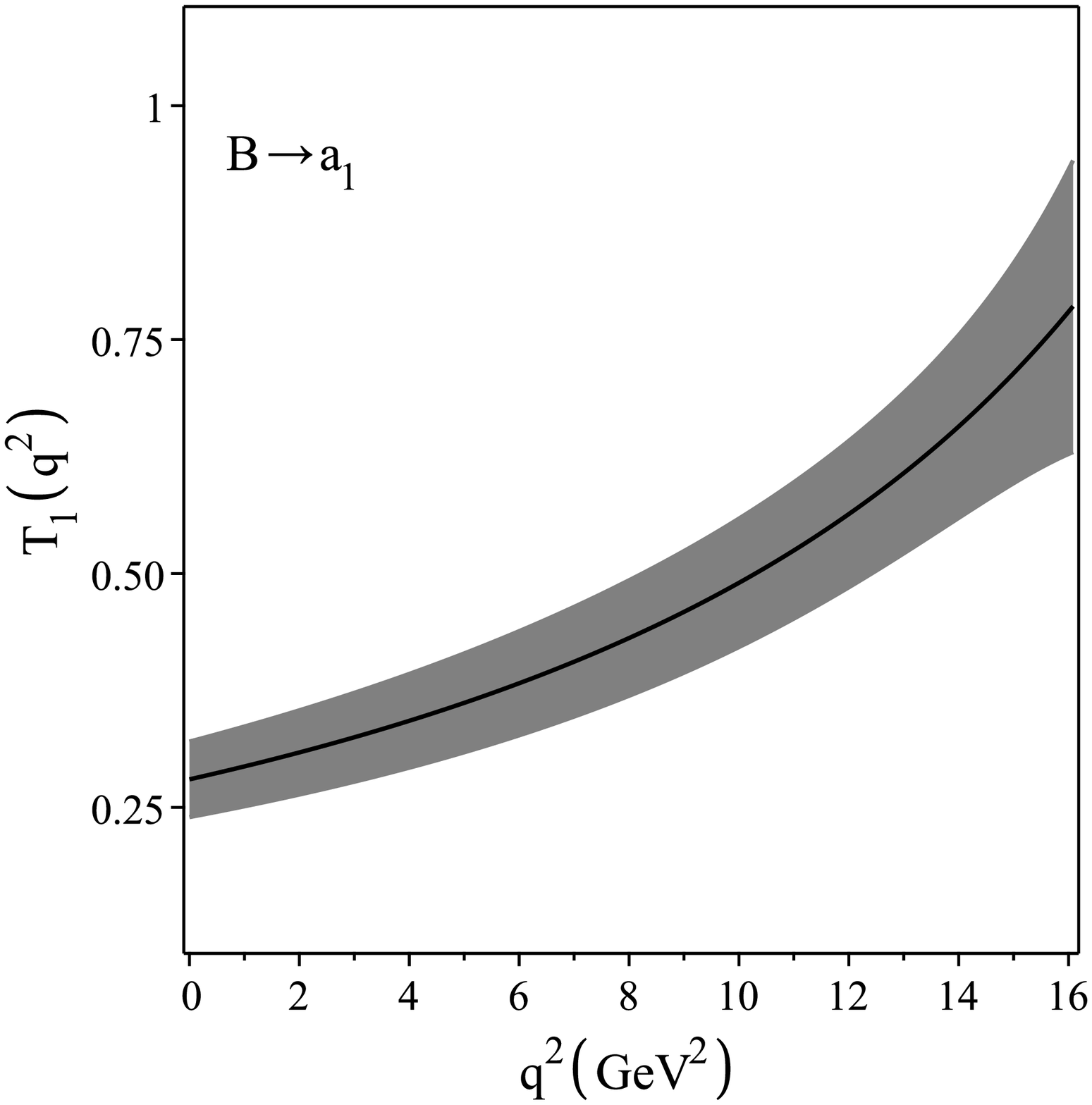}
\includegraphics[width=6cm,height=6cm]{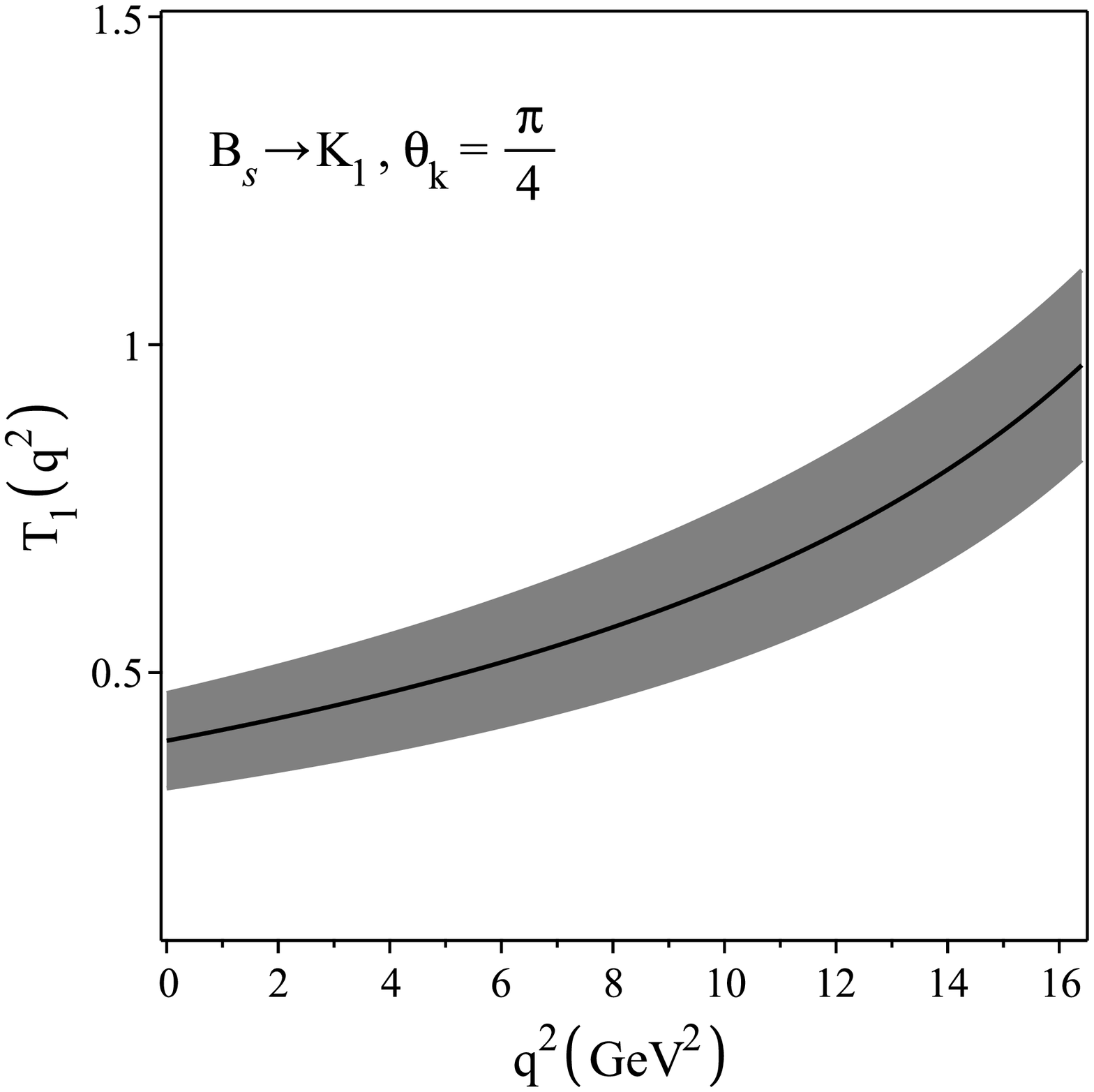}
\caption{The average form factors ${V}_{0}$ and ${T}_{1}$ of
$B_{(s)}\to a_1(K_1)$ decays with their uncertainty regions. }
\label{F43}
\end{figure}

By averaging the values of the form factors derived from the four
models of the $B$-meson DA's $\varphi_{\pm}$ at some points of
$q^2$, and then extrapolating to the fit function in Eq.
(\ref{eq31}), we can investigate  average form factors. The
parameters $F_{i}(0)$, $\alpha$ and $\beta$ for the average form
factors  of $B_{(s)} \to a_1(K_1)$ are given in Table \ref{T47}. For
$B_{s}\to K_1 \ell^{+} \ell^{-}$ transition, the average form
factors are calculated at $\theta_{K}=45^{\circ}$.
\begin{table}[th]
\caption{The parameters $F_{i}(0), \alpha$, and $\beta$ obtained for
the average form factors of the $B_{(s)}\to a_1(K_1)$ transitions. }
\label{T47}
\begin{ruledtabular}
\begin{tabular}{ccccc}
$\mbox{Form factor}$  &    $F_{i}(0)$   &    $\alpha$   &   $\beta$\\
\hline
$[A^{B\to a_1}, A^{B_s\to K_{1}}]$&$[0.26, 0.30 ]$&$[1.51, 1.17]$&$[0.53, 0.45]$\\
$[V_1^{B\to a_1}, V_1^{B_s\to K_{1}}]$&$[0.46, 0.48 ]$&$[1.05, 1.43]$&$[0.36, 0.16 ]$\\
$[V_2^{B\to a_1}, V_2^{B_s\to K_{1}}]$&$[0.23, 0.23 ]$&$[1.18, 1.08]$&$[0.55, 0.43] $\\
$[V_0^{B\to a_1}, V_0^{B_s\to K_{1}}]$&$[0.25, 0.15 ]$&$[1.17, 1.01]$&$[0.28, 0.30]$\\
$[T_1^{B\to a_1}, T_1^{B_s\to K_{1}}]$&$[0.28, 0.40]$&$[1.32, 1.18]$&$[0.36, 0.25]$\\
$[T_2^{B\to a_1}, T_2^{B_s\to K_{1}}]$&$[0.28, 0.40  ]$&$[1.29, 1.08]$&$[0.38, 0.27]$\\
$[T_3^{B\to a_1}, T_3^{B_s\to K_{1}}]$&$[0.24, 0.34]$&$[1.32, 0.86]$&$[0.83, 1.02]$\\
\end{tabular}
\end{ruledtabular}
\end{table}
Fig. \ref{F42} show the form factors with the four models, for
instance $V_{0}(q^2)$ and $T_{1}(q^2)$ with respect to $q^2$, on
which blue lines display the average  form factors.
Considering the uncertainties, the average form factors $V_{0}(q^2)$
and $T_{1}(q^2)$ of the $B_{(s)}\to a_1(K_1)$ decays with their
uncertainty regions are displayed on $q^2$ in Fig. \ref{F43}.

\begin{figure}
\includegraphics[width=6.5cm,height=5.5cm]{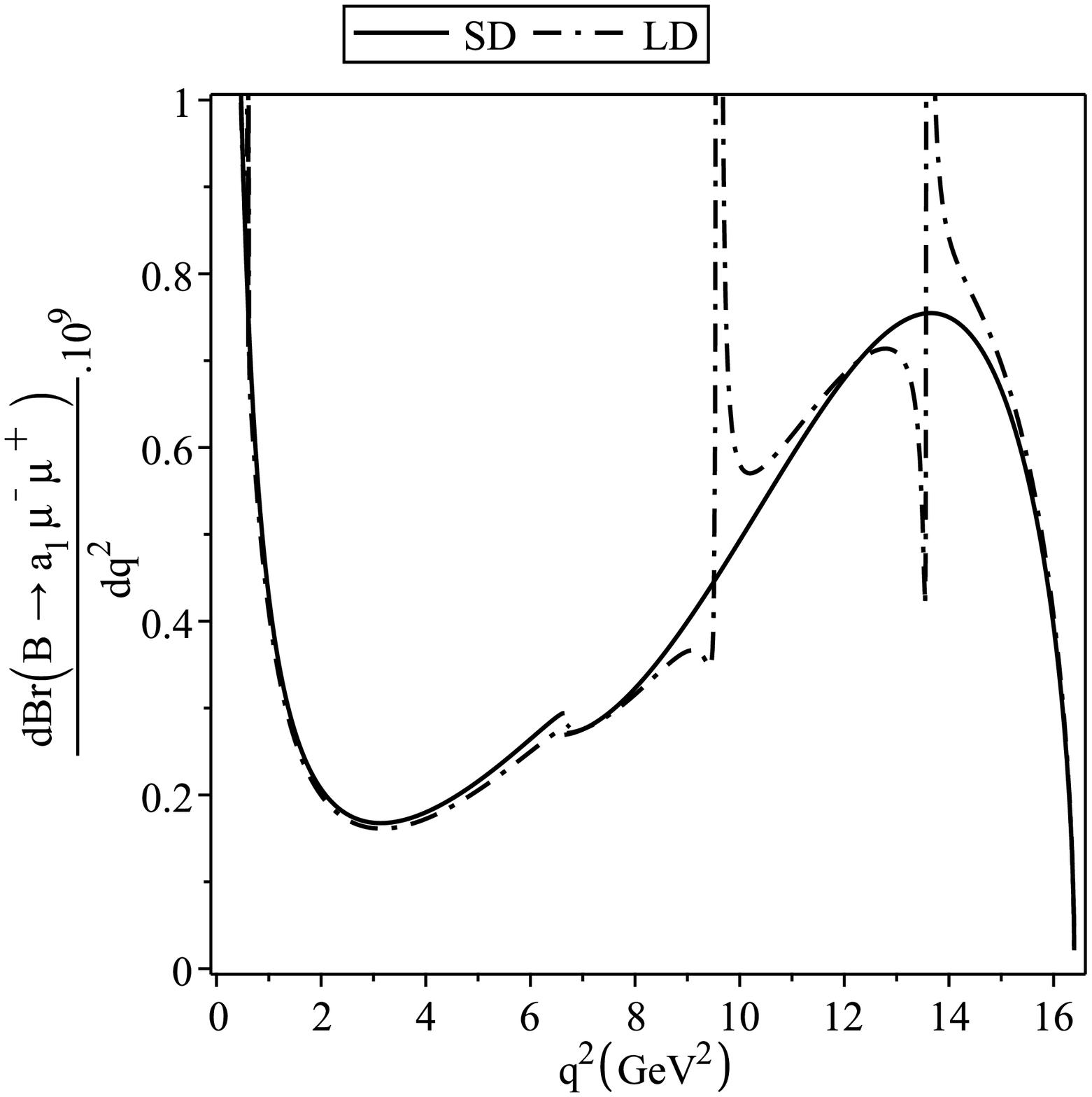}
\includegraphics[width=6.5cm,height=5.5cm]{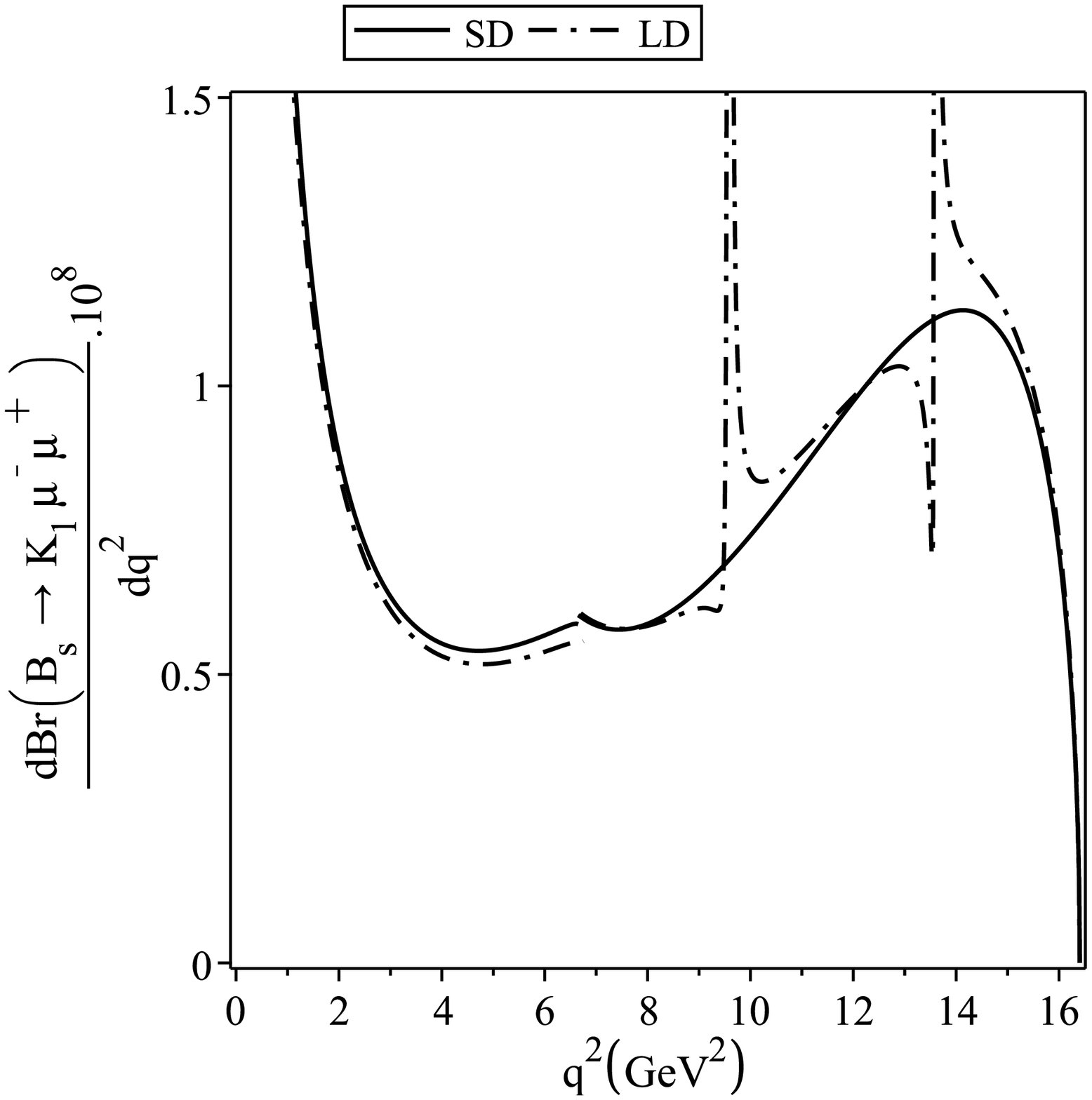}
\includegraphics[width=6.5cm,height=5.5cm]{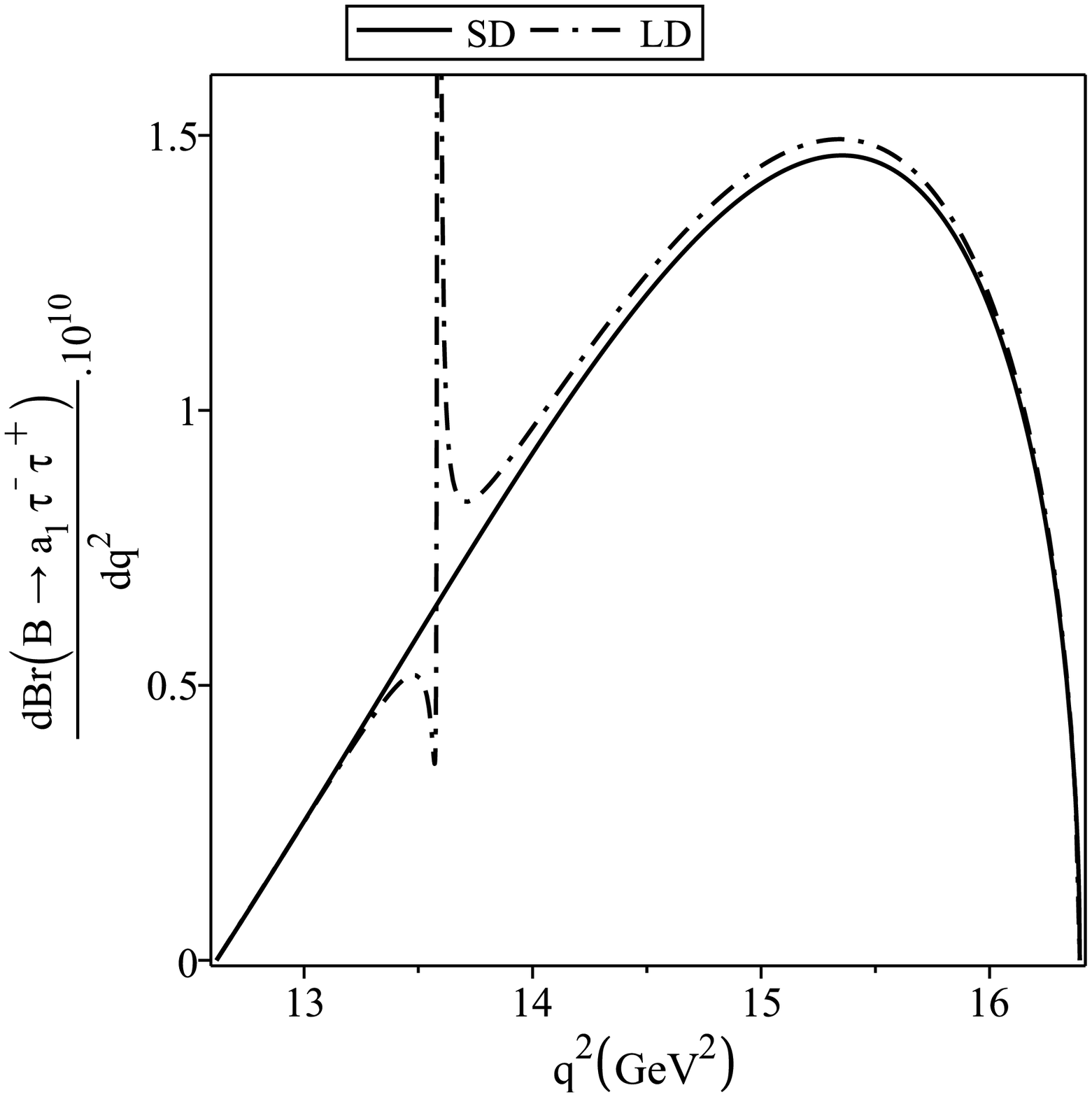}
\includegraphics[width=6.5cm,height=5.5cm]{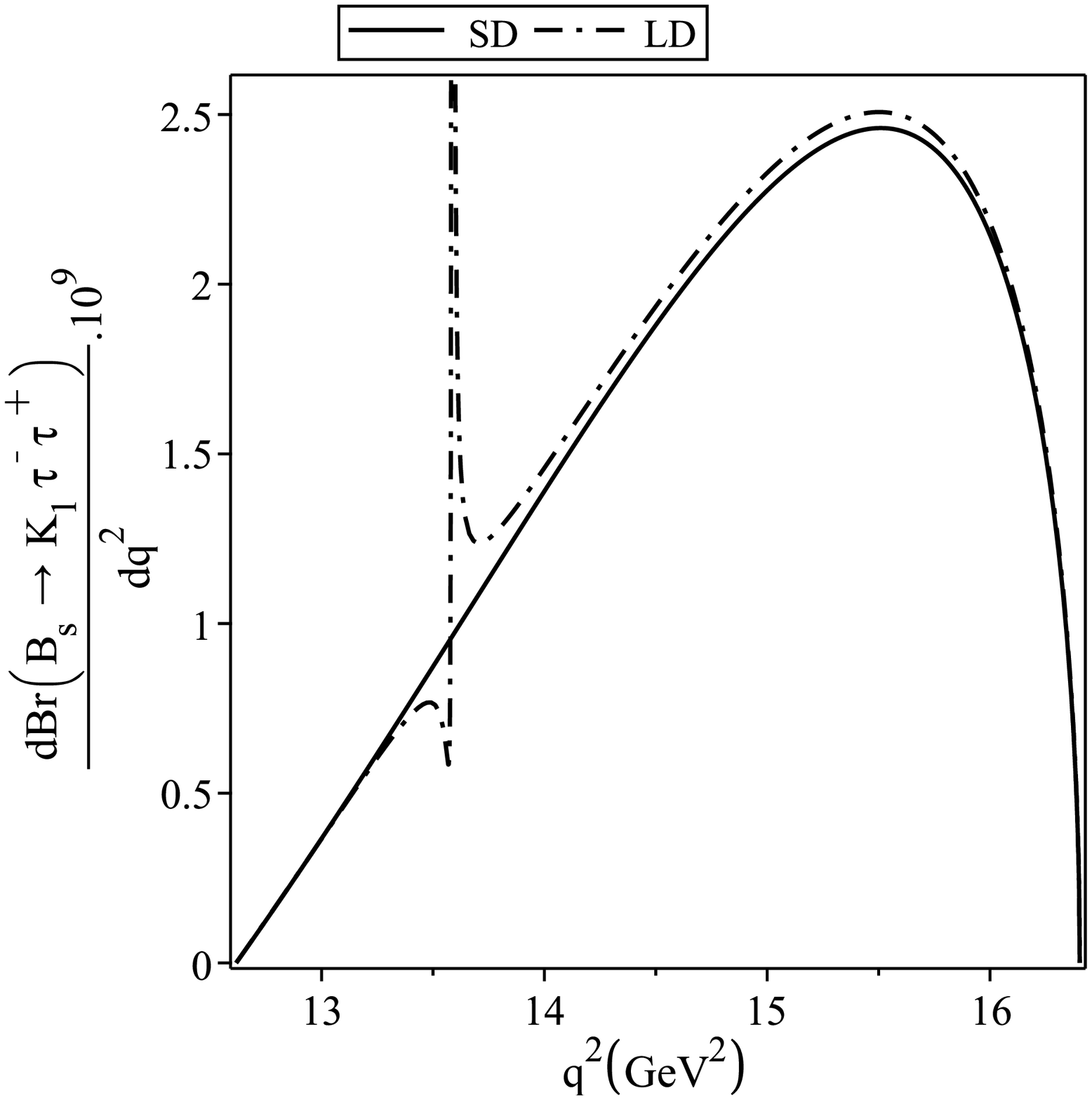}
\caption{The differential branching ratios of the semileptonic $B
\to a_1 (K_1) \ell^{+}\ell^{-}$ for $ \ell=\mu,\tau$ decays on $q^2$
with and without  LD effects using the average form factors.   }
\label{F44}
\end{figure}

Now, we are ready to evaluate the branching ratio values for the
$B_{(s)}\to a_{1}(K_1) \ell^{+} \ell^{-}$ decays. The expression of
double differential decay rate  ${d^2\Gamma}/{dq^2
dcos\theta_{\ell}} $ for the $ B_{(s)}\to a_{1}(K_1)$ transitions
can be found in \cite{Geng2, {Colangelo}}. This expression contains
the Wilson coefficients, the CKM matrix elements, the form factors
related to the fit functions, series of functions and  constants.
The numerical values of the Wilson coefficients are taken from Ref.
\cite{Ali}. The corresponding values are listed in Table \ref{Tw} in
the scale $\mu=m_b$.
\begin{table}
\caption{Central values of  the Wilson coefficients used in the numerical calculations.}
\begin{center}
\begin{tabular}{ccccccccc}
\hline
\hline
\ \ \ \ $C_1$ \ \ \ \ & \ \ \ \ $C_2$\ \ \ \  & \ \ \ \ $C_3$\ \ \ \  & \ \ \ \ $C_4$\ \ \ \  & \ \ \ \ $C_5$\ \ \ \ &
\ \ \ \ $C_6$ \ \ \ \ & \ \ \ \  $C_7^{\rm eff}$ \ \ \ \ & \ \ \ \ $C_9$ \ \ \ \ & \ \ \ \  $C_{10}$ \\
\hline
 -0.248 & 1.107&  0.011 & -0.026  & 0.007&  -0.031& -0.313&4.344& $-4.669$\\
\hline
\hline
\end{tabular}
\label{Tw}
\end{center}
\end{table}
The other parameters can be found in \cite{Colangelo}. After
numerical analysis, the dependency of the differential branching
ratios for $B \to a_1 \ell^{+}\ell^{-}$ on $q^2$ using the average
form factors, with and without LD effects is shown in Fig. \ref{F44}
for charged lepton case.

In Table \ref{T48}, we present the branching ratio values for muon
and tau without and with LD effects using the form factors derived
in the four models of $\varphi_{\pm}$. We also estimate  the
branching ratio values with the average form factors (AFF). For $B_s
\to K_1$ transitions, we have calculated the average value of
branching ratios in the region $35^\circ< |\theta_K| < 55^\circ$.
Here, we should also stress that the results obtained for the
electron are very close to those of the muon; and for this reason,
we only present the branching ratios for muon  in our table.
\begin{table}[th]
\caption{Branching ratio values of the semileptonic $B_{(s)} \to
a_1(K_1) \ell^{+}\ell^{-}$ decays without and with LD effects using
the form factors in the four models  as well as the average form
factors (AFF).} \label{T48}
\begin{ruledtabular}
\begin{tabular}{ccccccc}
\mbox{Only~ SD ~effects} & $\mbox{model-I}$ & $\mbox{model-II}$& $\mbox{model-III}$& $\mbox{model-VI}$&$\mbox{AFF}$\\
\hline
${\rm{BR}} (B \to a_1 \mu^+ \mu^-)\times 10^{8}  $& ${2.46}\pm {0.54}$ & ${2.77}\pm {0.58}$& ${2.96}\pm {0.62}$& ${2.43}\pm {0.50}$&${2.82} \pm {0.62}$\\
${\rm{BR}} (B \to K_1 \mu^+ \mu^-)\times 10^{8}  $& ${3.12}\pm{0.58}$ & ${3.41}\pm{0.82}$& ${3.64}\pm{0.94}$& ${3.09}\pm{0.57}$&${3.45} \pm {0.90}$\\
${\rm{BR}}(B \to a_1 \tau^+ \tau^-)\times10^{9} $& ${0.23}\pm{0.05}$ & ${0.25}\pm{0.05}$& ${0.29}\pm{0.06}$& ${0.22}\pm{0.04}$&
${0.27} \pm {0.06}$\\
${\rm{BR}}(B \to K_1 \tau^+ \tau^-)\times10^{9} $& ${0.40}\pm{0.09}$ & ${0.44}\pm{0.10}$& ${0.49}\pm{0.11}$& ${0.39}\pm{0.08}$&
${0.43 \pm 0.10}$\\
\hline
\mbox{SD + LD ~effects} &$\mbox{model-I}$ & $\mbox{model-II}$& $\mbox{model-III}$& $\mbox{model-VI}$&$\mbox{AFF}$\\
\hline
${\rm{BR}} (B \to a_1 \mu^+ \mu^-)\times 10^{8}  $&${2.82}\pm{0.65}$ & ${3.18}\pm{0.74}$& ${3.40}\pm{0.81}$& ${2.79}\pm{0.64}$&
${3.26} \pm {0.81}$\\
${\rm{BR}} (B \to K_1 \mu^+ \mu^-)\times 10^{8}  $&${3.80}\pm{0.83}$ & ${4.16}\pm{0.91}$& ${4.44}\pm{0.97}$& ${3.76}\pm{0.82}$&
${4.24} \pm {0.95}$\\
${\rm{BR}}(B \to a_1 \tau^+ \tau^-)\times10^{9} $& ${0.24}\pm{0.05}$ & ${0.26}\pm{0.05}$& ${0.31}\pm{0.06}$& ${0.23}\pm{0.04}$&
${0.29} \pm {0.06}$\\
${\rm{BR}}(B \to K_1 \tau^+ \tau^-)\times10^{9} $& ${0.45}\pm{0.10}$
& ${0.50}\pm{0.11}$& ${0.54}\pm{0.12}$& ${0.44}\pm{0.09}$&${0.49} \pm {0.11}$
\end{tabular}
\end{ruledtabular}
\end{table}
In Ref. \cite{mo} via the LCSR with $a_1$-meson DA's, the branching
ratio values of $B\to a_1 \mu^+ \mu^-$ and $B\to a_1 \tau^+ \tau^-$
decays by considering SD+ LD effects  are predicted $(2.52 \pm 0.62)
\times 10^{-8}$ and $(0.31 \pm 0.06) \times 10^{-9}$, respectively.
Our results are in a good agreement with its prediction for tau
case.

In summary, we calculated  the transition  form factors of the
$B_{(s)}\to a_{1}(K_1) \ell^{+} \ell^{-}$  decays via the LCSR with
the $B$-meson DA's, $\varphi_{\pm}$ in four models.  The main
uncertainty comes from the $\omega_{0}$ as a parameter of the
$B$-meson DA's. We estimated the branching ratio values for these
decays. The dependence of the differential branching ratios on $q^2$
were investigated. The results for branching fraction of $B\to a_1
\tau^+ \tau^-$ are in a good agreement with the usual LCSR method in
Ref. \cite{mo}. However, there is not good agreement between our
results for the form factors of $B\to a_1$ decays in $q^2=0$ with
those of the LCSR method.

\section*{Acknowledgments}
Partial support of the Isfahan university of technology research
council is appreciated.

\clearpage
\appendix
\begin{center}
{\Large \textbf{Appendix}}
\end{center}
In this appendix, the explicit expressions for  the  form factors of
$B \to a_1 \ell^+ \ell^-$ decays are given.
\begin{eqnarray*}\label{eq223}
V_{2}(q^{2})&=&\frac{f_{B}m_{B}(m_{B}-m_{a_1})}{f_{a}m_{a_1}} e^\frac{m_{a_1}^2}{M^2}\left\{\int_{0}^{\sigma_{0}}d\sigma \left[\frac{\sigma}{2\bar{\sigma}} \varphi_{_+}(\omega^{'}){e^{-\frac{s}{M^2}}}+\frac{1+\sigma}{2\bar{\sigma}} ( \widetilde{\varphi}_{_+}(\omega^{'})-\widetilde{\varphi}_{_-} (\omega^{'})){e^{-\frac{s}{M^2}}}\right] \right. \nonumber\\
&+&\left. \hat{\mathcal L} \left[\frac{(\sigma-2)(2u-1)}{2\bar{\sigma}^{2}{M^2}}(\Psi_{_A}-\Psi_{_V})~e^{-\frac{s}{M^2}}-\frac{(\sigma-2)(u-1)}{\bar{\sigma}^{2}{M^2}} \Psi_{_V}~ e^{-\frac{s}{M^2}}+\frac{6u-1}{2\bar{\sigma}^{2}M^2}\widetilde{X}_{_A}e^{-\frac{s}{M^2}}\right.\right.\nonumber\\
&-&\left.\left. \frac{\sigma(2u+1)-2}{2\bar{\sigma}^{2}M^2} \widetilde{X}_{_A}\frac{d}{d\sigma} e^{-\frac{s}{M^2}}-\frac{m_{B}^{2}(1+\sigma)(1-2u) }{\bar{\sigma}^{2} M^4}\widetilde{Y}_{_A}e^{-\frac{s}{M^2}}\right]\right\},
\end{eqnarray*}
\begin{eqnarray*}\label{eq224}
V_{0}(q^{2})-V_{3}(q^{2})&=&\frac{f_{B} m_{B} q^{2}}{f_{a}m_{a_1}^2}e^{\frac{m_{a_1}^{2}}{M^2}} \left\{\int_{0}^{\sigma_{0}}d\sigma\left[\frac{\sigma}{\bar{\sigma}} \varphi_{_+}({\omega^{'}}) {e^{-\frac{s}{M^2}}}
+\frac{\sigma^2}{\bar{\sigma}^2} (\widetilde{\varphi}_{_+}({\omega}^{'})-\widetilde{\varphi}_{_-} ({\omega}^{'})){e^{-\frac{s}{M^2}}}\right]\right.\nonumber\\
&+& \left.  \hat{\mathcal L} \left[\frac{\sigma(1-2u)}{\bar{\sigma}^2M^2}(\Psi_{_A}-\Psi_{_V}) e^{-\frac{s}{M^2}}+\frac{2\sigma(1-u)}{{\bar{\sigma}^2}{M^2}}\Psi_{_V}
{e^{-\frac{s}{M^2}}}-\frac{4u-2}{\bar{\sigma}^2 M^2 }\widetilde{X}{_{A}}e^{-\frac{s}{M^{2}}} \right. \right. \nonumber\\
&+&\left. \left.\frac{\sigma(1-2u)}{\bar{\sigma}^2 M^2 }\widetilde{X}{_{A}}\frac{d}{d\sigma}e^{-\frac{s}{M^{2}}}-
\frac{\sigma^{2} m_{B}^{2}(2-4u)}{\bar{\sigma}^3 M^4}\widetilde{Y}_{_A} e^{-\frac{s}{M^2}}\right]\right\},
\end{eqnarray*}
\begin{eqnarray*}\label{eq225}
A(q^{2})&=&\frac{2 f_{B}m_{B}}{{f_{a}m_{a_1}}(m_{B}-m_{a_1})}e^{\frac{m_{a_1}^2}{M^2}}\left \{\int_{0}^{\sigma_{0}}d\sigma \left[\frac{\sigma}{\bar{\sigma}}{\varphi_{_+}({\omega^{'}})}{e^{-\frac{s}{M^2}}}- \frac{\sigma}{\bar{\sigma}}(\widetilde{\varphi}_{_+}({\omega}')-\widetilde{\varphi}_{_-} ({\omega}')){e^{-\frac{s}{M^2}}}\right]\right.\nonumber\\
 &-&\left. \hat{\mathcal L}\left[ \frac{\sigma(1-2u)}{\bar{\sigma}^{2}M^{2}}(\Psi_{A}-\Psi_{V})e^{-\frac{s}{M^2}}
-\frac{2\sigma(1-u)}{\bar{\sigma}^{2}M^{2}}\Psi_{_V}e^{-\frac{s}{M^2}}-\frac{2\sigma m_{B}^{2}}{\bar{\sigma}^{2}{M^{4}}} (\widetilde{X}_{{A}}-\widetilde{Y}_{{A}}){e^{-\frac{s}{M^2}}}\right] \right\},
\end{eqnarray*}
\begin{eqnarray*}\label{eq226}
T_{1}(q^{2})&=&\frac{f_{B}m_{B}}{2m_{a_1}f_{a_1}}e^{\frac{m_{a_1}^{2}}{M^2}}\left\{ \int_{0}^{\sigma_{0}}d\sigma\left[\frac{\sigma}{\bar{\sigma}}\varphi_{_+}({\omega}^{'}){e^{-\frac{s}{M^2}}}+ \frac{1}{M^2}(\widetilde{\varphi}_{_+}({\omega}^{'})-\widetilde{\varphi}_{_-} ({\omega}^{'}))\frac{d}{d\sigma}e^{-\frac{s}{M^2}}\right.\right. \nonumber\\
&+&\left.\left. \frac{\mathcal M^{2}(\bar\sigma+\sigma ^{2})+\bar{\sigma}+1}{\bar{\sigma}^{2} M^{2}}(\widetilde{\varphi}_{_+}({\omega}^{'})-\widetilde{\varphi}_{_-} ({\omega}^{'}))e^{-\frac{s}{M^2}}\right]+ \hat{\mathcal L}\left[\frac{\sigma ^{2}(1-u)}{2\bar{\sigma}^{2}M^2}{\Psi_{_V}}e^{-\frac{s}{M^2}}\right.\right. \nonumber\\
&+&\left. \left.\frac{\sigma ^{2}(1-2u)}{\bar{\sigma}^{2}M^2}(\Psi_{_A}-\Psi_{_V})e^{-\frac{s}{M^2}}+\frac{\mathcal M^{2}(\bar{\sigma}^{2}(\sigma-1)+2\bar{\sigma}-\sigma)}{\bar{\sigma}M^2}(\widetilde{X}_{_A}-\widetilde{Y}_{_A})e^{-\frac{s}{M^2}} \right. \right. \nonumber\\
&-&\left.\left.\left(\frac{3\bar{\sigma}^{2}-2\sigma^{2}}{\bar{\sigma}^{2}}+\frac{2\mathcal M^{2}u}{M^{4}}-\frac{2(4\bar{\sigma}+3)u}{4\bar{\sigma}M^2}\right) (\widetilde{X}_{_A}-\widetilde{Y}_{_A})e^{-\frac{s}{M^2}}-\frac{\mathcal M^{2}u}{M^{4}}\widetilde{X}_{_A}e^{-\frac{s}{M^2}}\right.\right. \nonumber\\
&+&\left.\left. \frac{(2\bar{\sigma}+3)u}{2\bar{\sigma}M^2} \widetilde{X}_{_A}e^{-\frac{s}{M^2}}\right]\right\},
\end{eqnarray*}
\begin{eqnarray*}\label{eq227}
T_{2}(q^{2})&=&\frac{f_{B}m_{B}}{2f_{a}m_{a_1}(m_{B}^{2}-m_{a_1}^{2})} e^{\frac{m_{a_1}^2}{M^2}} \left\{\int_{0}^{\sigma_{0}}d\sigma\left[(s+\sigma m_{B}^{2}){\varphi_{_+}({\omega}^{'})}{e^{-\frac{s}{M^2}}}
+\left(\frac{\mathcal M^{'2}}{2}+\frac{\mathcal M^{4}}{M^{2}}\right) \right.\right. \nonumber\\
&\times & \left. \left. (\widetilde{\varphi}_{_+}({\omega}')-\widetilde{\varphi}_{_-} ({\omega}')) {e^{-\frac{s}{M^2}}}\right]
+ \hat{\mathcal L}\left[(1-2u)\Psi_{_A}\frac{d}{d\sigma}{e^{-\frac{s}{M^2}}}+(4u-3)\Psi_{_V}\frac{d}{d\sigma}{e^{-\frac{s}{M^2}}}\right.\right. \nonumber\\
&+&\left.\left. (4u-2)  \left(\frac{3}{\bar{\sigma}^{2}}+\frac{\mathcal M^{'2}}{\bar{\sigma}^{2} M^2}-\frac{\mathcal M^{4}}{2\bar{\sigma}^{2} M^{4}}\right)(\widetilde{X}_{{A}}-\widetilde{Y}{_{A}})e^{-\frac{s}{M^2}}
+ (4u-2) \frac{\mathcal M^{2}}{\bar{\sigma}^{2} M^2}\widetilde{Y}_{{A}} e^{-\frac{s}{M^2}}\right]\right\},
\end{eqnarray*}
\begin{eqnarray*}\label{eq228}
T_{3}(q^{2})&=&\frac{f_{B}m_{B}}{f_{a}m_{a_1}}e^{\frac{m_{a_1}^2}{M^2}}\left \{\int_{0}^{\sigma_{0}}d\sigma\left[ \frac{\sigma}{\bar{\sigma}} {\varphi_{_+}({\omega}')} e^{-\frac{s}{M^2}}
-\left(\frac{4}{M^{2}}-\frac{4\sigma\mathcal M^{2}}{\bar{\sigma}M^{4}}-
\frac{2\sigma^{2}(\mathcal M^{2}+\mathcal M^{'2})}{\bar{\sigma} M^{4}}\right) \right.\right. \nonumber\\
&\times&\left.\left.   (\widetilde{\varphi}_{_+}({\omega}^{'})-\widetilde{\varphi}_{_-} ({\omega}^{'}))\frac{{e^{-\frac{s}{M^2}}}}{\bar{\sigma}}\right]+ \hat{\mathcal L}\left[ (1-2u)\left(\frac{1}{M^2}+\frac{\mathcal M^{2}}{M^{4}}-\frac{m_{B}^{2}}{M^{4}}\right) (\Psi_{_A}-\Psi_{_V})e^{-\frac{s}{M^2}}\right.\right. \nonumber\\
&+&\left.\left. \frac{\sigma (u-1) }{\bar{\sigma}^{2}M^2}{\Psi_{_V}}
e^{-\frac{s}{M^2}}+(1-2u)\left(
\frac{1}{\bar{\sigma}^2M^2}+\frac{2\sigma}{\bar{\sigma}^3
M^2}-\frac{\sigma\mathcal M^{2}}{\bar{\sigma}^3M^4}\right)
(\widetilde{X}_{{A}}-2\widetilde{Y}_{{A}}) e^{-\frac{s}{M^2}}\right]
\right\},
\end{eqnarray*}
where:
\begin{eqnarray*}\label{eq229}
\hat{\mathcal L}\equiv\int_{0}^{\sigma_{0}}d\sigma
\int_{0}^{\omega'}d\omega \int_{\omega'-\omega}^{\infty
}\frac{d\xi}{\xi},
\end{eqnarray*}
$\omega^{'}=\sigma m_{B}$ and
$\widetilde{m}_{B}^2=m_B^2(1+\sigma)-{q^2}/{\bar{\sigma}}$,
 also:
\begin{eqnarray*}\label{eq230}
&&\begin{array}{lll} s                      =     \sigma m_B^2
-\frac{\sigma}{\bar{\sigma}}q^2, ~~~~~~~~~~~~~~~~~~~~~~~~&
\bar{\sigma} = 1-\sigma, ~~~~~~~~~~~~~~~~~~~~~~~~&
u                       =     \frac{\omega'-\omega}{\xi},                                        \\
\widetilde{\phi}_{_\pm} =    \int_0^\omega d\tau \phi_\pm(\tau), &
\widetilde{X}{_A}       =     \int_0^\omega d\tau X_{_A}(\tau,\xi),
&
\widetilde{Y}{_A}       =     \int_0^\omega d\tau Y_{_A}(\tau,\xi),                                   \\
\mathcal M^{2}          =     \widetilde{m}_{B}-2\sigma m_{B}^{2},
&\mathcal M^{4}          = \widetilde{m}_{B}^{4}-4s~m_{B}^{2}, &
\mathcal M^{'2}         = 2m_{B}^{2}-\widetilde{m}_{B}^{2},
\end{array}\nonumber\\
&&\sigma_0=\frac{s_0+m_B^2-q^2-\sqrt{(s_0+m_B^2-q^2)^2-4s_0m_B^2}}{2m_B^2}.
\end{eqnarray*}


\begin{thebibliography}{II}

\bibitem{Kolesnichenko}
I. I. Balitsky, V. M. Braun and A. V. Kolesnichenko, Nucl. Phys.   B
{\bf 312}, 509 (1989).

\bibitem{Halperin}
V. M. Braun and I. E. Halperin, Z. Phys. C {\bf  44}, 157 (1989).

\bibitem{Zhitnitsky}
V. L. Chernyak and I. R. Zhitnitsky, Nucl. Phys. B {\bf  345}, 137
(1990).

\bibitem{Ruckl}
V. M. Belyaev, A. Khodjamirian and R. Ruckl, Z. Phys. C {\bf  60},
349 (1993).

\bibitem{Simma}
A. Ali, V. M. Braun and H. Simma, Z. Phys. C {\bf  63}, 437 (1994).

\bibitem{Belyaev}
V. M. Belyaev, V. M. Braun, A. Khodjamirian and R. Ruckl, Phys. Rev.
D {\bf  51}, 6177 (1995).

\bibitem{Weinzierl}
A. Khodjamirian, R. Ruckl, S. Weinzierl and O. I. Yakovlev, Phys.
Lett. B {\bf  410}, 275 (1997).

\bibitem{BBraun}
P. Ball and V. M. Braun, Phys. Rev. D {\bf  58}, 094016 (1998).

\bibitem{Bagan}
E. Bagan, P. Ball and V. M. Braun, Phys. Lett. B {\bf  417}, 154
(1998).

\bibitem{Ball}
P. Ball, JHEP {\bf 9809}, 005 (1998).

\bibitem{Zwicky}
P. Ball and R. Zwicky, JHEP {\bf 0110}, 019 (2001).

\bibitem{BZwicky}
P. Ball and R. Zwicky, Phys. Rev. D {\bf  71}, 014015 (2005).

\bibitem{Brodsky}
A. Szczepaniak, E. M. Henley and S. J. Brodsky, Phys. Lett. B {\bf
243},  287 (1990).

\bibitem{Grozin}
A. G. Grozin and M. Neubert, Phys. Rev. D {\bf  55}, 272 (1997).

\bibitem{Beneke}
M. Beneke and T. Feldmann, Nucl. Phys.  B {\bf  592}, 3 (2001).

\bibitem{Kou}
P. Ball and E. Kou, JHEP {\bf 0304}, 029 (2003).

\bibitem{Offen}
A. Khodjamirian, T. Mannel and N. Offen,  Phys. Lett. B {\bf  620},
52 (2005).

\bibitem{Mannel}
A. Khodjamirian, T. Mannel and N. Offen, Phys. Rev. D {\bf  75},
054013 (2007).

\bibitem{Hatanaka}
H. Hatanaka and K. C. Yang, Phys. Rev.  D {\bf 78}, 074007 (2008).

\bibitem{Burakovsky}
L. Burakovsky and T. Goldman, Phys. Rev. D {\bf 57}, 2879 (1998).

\bibitem{Suzuki}
M. Suzuki, Phys. Rev. D {\bf 47}, (1993) 1252.

\bibitem{Buras0}
A. J. Buras and M. Muenz, Phys. Rev. D {\bf 52}, 186 (1995).

\bibitem{Aliev0}
T. M. Aliev, V. Bashiry and M. Savci, Phys, Rev, D {\bf 72}, 034031
(2005).

\bibitem{Jezabek}
M. Jezabek and J. H. Kuhn, Nucl. Phys. B {\bf 320}, 20 (1989).

\bibitem{Misiak}
M. Misiak, Nucl. Phys. B {\bf 439}, 461 (1995).

\bibitem{Colangelo}
P. Colangelo, F. De Fazio, P. Santorelli and  E. Scrimieri, Phys.
Rev. D {\bf   53}, 3672 (1996).


\bibitem{Wang}
Z. G .Wang ,Phys, Lett. B {\bf  666}, 477-482 (2008).

\bibitem{Braun}
V. M. Braun, D. Yu. Ivanov and G. P. Korchemsky, Phys. Rev. D {\bf
69}, 034014 (2004).

\bibitem{Faziophi}
F. De Fazio, T. Feldmann and T. Hurth, JHEP {\bf 0802}, 031 (2008).

\bibitem{Wangphi}
Y. M. Wang and Y. L. Shen, Nucl. Phys. B {\bf 898}, 563 (2015).

\bibitem{Hua}
Y. Li, J. Hua and  K. Yang, Eur. Phys. J. C {\bf 71}, 1775 (2011).

\bibitem{pdg}
 K. A. Olive et al., Particle Data Group, Chin. Phys. C  {\bf 38}, 090001 (2014).

\bibitem{Yang}
K. C. Yang, Nucl. Phys. B {\bf   776}, 187 (2007).

\bibitem{Wan}
Z. G. Wang, W. M. Yang and S. L. Wan, Nucl. Phys. A {\bf  744}, 156
(2004).

\bibitem{Li2}
R. Li, C. Lu and W. Wang, Phys. Rev. D {\bf  79}, 034014 (2009).

\bibitem{mo}
S. Momeni, R. Khosravi, and F. Falahati,  Phys. Rev. D {\bf  95}, 016009 (2017).

\bibitem{kh}
R. Khosravi, Eur. Phys. J. C {\bf 75}, 220 (2015).

\bibitem{Geng2}
C. Q. Geng and C. C. Liu, J. Phys. G {\bf  29}, 1103 (2003).

\bibitem{Ali}
A. Ali, P. Ball, L. T. Handoko and G. Hiller, Phys. Rev.  D {\bf 61}
074024 (2000).





\end{thebibliography}
\end{document}